\documentclass[12pt]{article}

\usepackage{a4}
\usepackage[english]{babel}
\usepackage{amsfonts}
\usepackage{pslatex}
\usepackage[latin1]{inputenc}
\usepackage[T1]{fontenc}

\ifx\pdfoutput\undefined
\usepackage{hyperref}
\usepackage{graphicx}
\else
\usepackage[pdftex]{hyperref}
\usepackage[pdftex]{graphicx}
\fi

\makeatletter
\def\@sect#1#2#3#4#5#6[#7]#8{\ifnum #2>\c@secnumdepth
  \def\@svsec{}\else 
  \refstepcounter{#1}\edef\@svsec{\csname the#1\endcsname.\hskip0.5em}\fi
  \@tempskipa #5\relax
  \ifdim \@tempskipa>\z@
    \begingroup 
      #6\relax
      \@hangfrom{\hskip #3\relax\@svsec}{\interlinepenalty \@M #8\par}%
    \endgroup
    \csname #1mark\endcsname{#7}\addcontentsline
      {toc}{#1}{\ifnum #2>\c@secnumdepth \else
        \protect\numberline{\csname the#1\endcsname}\fi #7}%
  \else
    \def\@svsechd{#6\hskip #3\@svsec #8\csname #1mark\endcsname
      {#7}\addcontentsline{toc}{#1}{\ifnum #2>\c@secnumdepth \else
        \protect\numberline{\csname the#1\endcsname}\fi #7}}%
  \fi \@xsect{#5}}
\@addtoreset{equation}{section}
\@addtoreset{figure}{section}
\makeatother

\renewcommand\theequation{\ifnum \value{section}>0
 \arabic{section}.\arabic{equation}%
\else
\arabic{equation}%
\fi}
\renewcommand\thefigure{\ifnum \value{section}>0
 \arabic{section}.\arabic{figure}%
\else
\arabic{figure}%
\fi}

\def\ksl{k\!\!\!/}
\def\bra{\langle}
\def\ket{\rangle}
\def\st#1{{\mbox{\scriptsize #1}}}
\def\on{{\st{on}}}
\def\cN{{\cal N}}

\def\cM{{\cal M}}
\def\cS{{\cal S}}
\def\L{\left(}
\def\R{\right)}
\def\kt{{k_t}}
\def\kw{{k_w}}
\def\kg{{k_\gamma}}
\def\kb{{k_b}}
\def\k1{{k_1}}
\def\xt{{x_t}}
\def\xb{{x_b}}
\def\zt{{z_t}}
\def\mb{{m_b}}
\def\mw{{m_w}}
\def\mt{{m_t}}

\def\zb{{z_b}}
\def\zw{{z_w}}

\def\x3{{x_3}}
\def\Gf{{G_f}}
\def\Cf{{C_F}}
\def\Vtb#1{{|V_{tb}|^{#1}}}
\def\cz{{z}}
\def\sigtot{\sigma_{\rm tot}}
\def\coll{\st{coll}}
\def\eik{\st{eik}}
\def\ltb{\lambda_{bt}}

\def\msbar{{\ensuremath{\overline{\mbox{MS}}}}}
\def\as{{\alpha_s}}
\def\Re{{\rm Re}}

\def\e{{\epsilon}}
\def\epsIR{{\e_{\rm IR}}}

\def\wt{{\omega_t}}
\def\wb{{\omega_b}}
\def\Li2{{{\rm Li}_2}}

\def\eq#1{{eq.~(\ref{#1})}}
\def\fig#1{{fig.~\ref{#1}}}
\def\PolSum{{\mbox{\st{$\gamma,t,b$ Pol.}}}}
\def\aem{\alpha}
\def\nn{\nonumber}

\def\Cf{C_F}

\begin{document}
\thispagestyle{empty}
\hfill  TTP03-11 
\vspace*{3cm}
\begin{center}
  {\Large\bf QCD corrections to 
    single top quark production\\ in electron photon interactions}\\
  \vspace*{1cm}
  J.H.~Kühn, C.~Sturm, and P.~Uwer\\
  \vspace*{0.5cm}
  {\em Institut für Theoretische Teilchenphysik,
    Universität Karlsruhe\\ 76128 Karlsruhe, Germany}
\end{center}
\vspace*{1.5cm}
\centerline{\bf Abstract}
\begin{center}
  \parbox{0.8\textwidth}{%
    Single top quark production in electron photon interactions provides a
    clean environment for the measurement of the 
    Cabbibo-Kobayashi-Maskawa matrix element $V_{tb}$. Aiming an 
    experimental precision at the percent level the knowledge of radiative 
    corrections is important. In this paper we present results for 
    the radiative corrections in quantum chromodynamics.}
\end{center}

\newpage
\setcounter{page}{1}
\section{Introduction}
\label{sec:intro}
One of the fundamental unsolved problems of todays high-energy physics is the
exact mechanism of electroweak symmetry breaking (EWSB). Due to the fact
that only the top quark couples with a Yukawa coupling of order one to
the so far unobserved Higgs boson it is natural to assume that 
the top quark plays a special rôle in the context of the EWSB. 
For example in so-called dynamical symmetry breaking models the scalar
Higgs field --- responsible for the spontaneous symmetry breaking in the 
Standard Model --- is replaced by a composite scalar operator.  Such an 
operator could be built for example from heavy fermion fields. 
Having eliminated
the elementary scalar field from the theory the problem of the 
large mass corrections due to quantum corrections is solved. Examples for such
models are technicolour models (for a review see  
ref. \cite{King:1990ec} and references therein), 
top-condensate models \cite{Bardeen:1990ds},
and top-colour models \cite{Hill:1991at,Hill:1995hp}. 
For the search of such extensions a 
precise understanding of the top quark sector of the standard model
is necessary. 

At hadron colliders both single top quark production as well as top
quark pair production have been studied extensively in the past. The
differential cross section for top quark pair production is known to 
next-to-leading order (NLO) accuracy in quantum chromodynamics (QCD) 
\cite{Nason:1988xz,Nason:1989zy,Beenakker:1989bq,Beenakker:1991ma}.
In addition the resummation of logarithmic enhanced contributions has
been studied in detail in refs.
\cite{Laenen:1992af,Kidonakis:1995wz,Berger:1996ad,Catani:1996yz,Berger:1998gz}.
Recently also the spin correlations between top quark and anti-top quark
were calculated at NLO in QCD \cite{Bernreuther:2001rq}.
Due to the fact that single top quark production provides an excellent 
opportunity to test the charged-current weak-interaction of the top
quark it has also attracted a lot of interest in the past. In
particular NLO corrections were studied in 
refs. \cite{Bordes:1995ki,Smith:1996ij,Stelzer:1997ns,Harris:2002md}. In 
ref. \cite{Harris:2002md} the NLO corrections for the fully
differential cross section is given keeping also the spin information 
of the top quark. On the experimental side the situation is not very
conclusive for the moment as far as single top quark production is 
concerned. Due to limited statistics at run I of the
Tevatron collider only upper bounds were obained in ref. \cite{Acosta:2001un}.
In ref. \cite{Espriu:2001vj} the possibility to measure the
electroweak couplings in single top quark production at the LHC is
studied. In particular also the sensitivity to new physics is
discussed.

As far as lepton colliders are concerned much effort has been devoted 
to top quark pair production in $e^+e^-$-annihilation. In particular
the total cross section in the threshold region is known at 
next-to-next-leading order (NNLO) in QCD. (For an overview of the
theoretical status we refer to ref. \cite{Hoang:2000yr}.)
Momentum \cite{Jezabek:1992np,Sumino:1993ai} as well as angular
\cite{Murayama:1993mg,Harlander:1995ac} distributions were studied in
detail.
In the continuum region the total cross section for massive quarks is known to 
order $\as^2$ in the coupling constant of the strong interaction
\cite{Chetyrkin:1996ii}. 
In order $\as^3$ the quartic mass corrections to the total cross
section are also known \cite{Chetyrkin:2000zk}. 
Also less inclusive observables have been studied in great detail. 
For example the spin structure of top anti-top system is 
completely known
at order $\as$ \cite{Groote:1996yc,Groote:1996ky,Groote:1997nc,Brandenburg:1998xw,Schmidt:1996mr} and partially known at order
$\as^2$ \cite{Ravindran:2000rz}. 
Futhermore the 3-jet results obtained for massive $b$-quarks
\cite{BeBrUw97a,BiRoSa97b,BrUw97,BiRoSa99,NaOl97a,NaOl97b} are also
applicable to top quark physics \cite{Brandenburg:1999fv}.

Less attention has been devoted to single top quark production.
Studies at tree level can be found for example in refs.
\cite{Panella:1993zk,Raidal:1993na,Ambrosanio:1994eu,Dokholian:1994nv,Boos:2001sj}.
In the refs. \cite{Panella:1993zk,Raidal:1993na} special emphasis was
put on single top quark production at LEP II. For a top quark
mass arround 175 GeV
the production rates in the standard model are to small to be
detectable at LEP II. This was recently confirmed by the L3
collaboration \cite{Achard:2002vv}. In ref. \cite{Boos:2001sj} also 
the single top quark production in electron photon collisions is studied.
The electron photon reaction provides a clean environment for the 
study of single top quark  production because there is no background from
top quark pair production.  As a consequence this reaction is very
well suited for the measurement of the weak couplings of the top quark.
In particular it has been shown in ref. \cite{Boos:2001sj} that 
using polarised beams the Cabibbo-Kobayashi-Maskawa (CKM) matrix 
element $V_{tb}$ can be
measured with an uncertainty of 1\% at the 2$\sigma$ level. In this
analysis $10^4$ top quark events were assumed which corresponds to a luminosity
of 100 fb$^{-1}$  of an electron photon collider operating 
at $\sqrt{s}=500$ GeV.\\ 
Aiming an accuracy at the percent level it is clear that the knowledge 
of the QCD corrections is mandatory. This is the main purpose of the
present paper. In addition we study the structure of logarithmic
enhanced contributions which are related to initial state
singularities. The full dependence on the $b$-quark mass is kept. This
allows a systematic comparison between the structure function approach
and the fixed-order calculation. Furthermore close to the threshold 
effects of the finite $b$-quark mass are important. To calculate
the QCD corrections we use the {\it effective W-approximation}. In
the $W$-approximation the scattering process which needs to be studied
is $W^+\gamma\to t\bar b$. Using the $W$-approximation which describes
the momentum distribution of the $W$-boson in the electron a
prediction for $e^+\gamma \to t\bar b \nu_e$ can be obtained. A more 
detailed discussion is given in  section \ref{sec:results}.

The outline of the paper is as follows. In section
\ref{sec:kinematics} we discuss the kinematics and the leading order
results for the reaction $W^+\gamma \to t\bar b$. The virtual
corrections to this process are discussed in section
\ref{sec:virtual}. In section \ref{sec:real} the real corrections are 
calculated. In particular the cancellation of the infrared
singularities is shown. In the following section we present the
results for the subreaction $W^+\gamma \to t\bar b$ as well as for 
the reaction $e^+\gamma\to t\bar b\bar \nu_e$. We finally close with our 
conclusions in section \ref{sec:conclusions}.

\section{Kinematics and leading-order results}
\label{sec:kinematics}
\begin{figure}[htbp]
  \begin{center}
    \includegraphics*[width=13cm,bb=29 577 569 667]{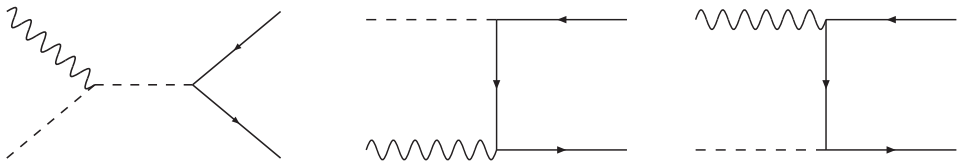}
    \caption{ Feynman diagrams for $W^+ \gamma \to t \bar b$ in
      leading-order%
      \label{fig:lo}}
  \end{center}
\end{figure}
In the following we study the reaction
\begin{equation}
  W^+(\kw) + \gamma(\kg) \to t(\kt) + \bar b(\kb),
  \label{eq:lo-reaction}
\end{equation}
where we treat both outgoing quarks as massive. For later use it
is convenient to define dimensionless variables. In particular we
define the rescaled masses
\begin{equation}
  z_i = {m_i^2\over s}
\end{equation}
and the energy fractions
\begin{equation}
  x_i = {2(k\cdot k_i)\over s} 
\end{equation}
with $k= \kw+\kg$ and $s=k^2$.
For the reaction given in \eq{eq:lo-reaction} the energy fractions
are fixed completely by the kinematics:
\begin{equation}
  \xt = 1 + \zt - \zb, \quad\mbox{and}\quad \xb = 1+\zb -\zt.
\end{equation}
This is no longer true when the emission of an additional 
gluon is considered (c.f. section \ref{sec:real} ). 
The calculation of the Born matrix elements for 
the reaction  (\ref{eq:lo-reaction}) is straightforward so we just
quote the results here. The corresponding Feynman diagrams are shown
in fig.~\ref{fig:lo}. 
For the $W$ boson we distinguish between
longitudinal and transverse polarization while for the photon we
average over the incoming polarization. For partons in the final state
the polarization (and the colour) is summed over.  
In terms of the leading-order squared matrix elements $|\cM_0^{T,L}|^2$ for
transversely/longitudinally polarized $W$-bosons
the differential cross sections are given by
\begin{equation}
  {d\sigma^{T,L}\over d\Omega}
  =
  {N\*\ltb\over 64\*\pi^2\*(1-\zw)\*s} \*{1\over \cN_{\,T,L}}
  \*\sum_{\PolSum}
  |\cM_0^{T,L}|^2,
\end{equation}
where $\ltb$ is defined as
\begin{equation}
  \ltb=\lambda(1,\zb,\zt)
\end{equation}
with 
\begin{equation}
  \lambda(x,y,z) = \sqrt{x^2+y^2+z^2-2\*x\*y-2\*x\*z-2\*y\*z}.
\end{equation}
The number of colours is denoted by $N$.
For transversely polarized $W$-bosons the normalization $\cN_T$ is
given by  $\cN_T=2 \cdot 2$. For 
longitudinally polarized $W$-bosons we have $\cN_L=2$. 
The squared matrix element for longitudinally polarized $W$-bosons 
is given by 
\begin{eqnarray}
  \sum_{\PolSum}|\cM_0^L|^2
  &=&{2\*\kappa\over9\*\zw\*(1-\zw)^2}
  \*\Bigg({8\*(\zb^3-2\*\zb^2\*\zt+\zb\*\zt^2+(3\*\zb^2-\zb\*\zt)\*\zw)
    \over\left(\xb-\ltb\*\cz\right)^2}
  \nn\\
  &+&
  {32\*(\zb^2\*\zt-2\*\zb\*\zt^2+\zt^3+(3\*\zt^2-\zb\*\zt)\*\zw)
    \over\left(\xt+\ltb\*\cz\right)^2}
  +{l_1^L\over\xb-\ltb\*\cz}
  +{l_2^L\over\xt+\ltb\*\cz}
  \nn\\
  &+&12\*\cz\*\ltb\*(3\*\zb-3\*\zt-\zw)\*\zw
  +18\*\cz^2\*\ltb^2\*\zw^2
  +18\*\zb^2-9\*\zb\*(1+4\*\zt)-9\*\zt
  \nn\\
  &+&18\*\zt^2 +\zw\*(36\*\zb^2+12\*\zb\*(1-6\*\zt)+60\*\zt+36\*\zt^2)
  \nn\\
  &+&
  \zw^2\*(2-18\*\zb^2+39\*\zt-18\*\zt^2+3\*\zb\*(5+12\*\zt))
  \Bigg),
\end{eqnarray}
where $\cz$ denotes the cosine of the scattering angle
\begin{equation}
  \cz=\cos\theta_{b\gamma}=-\cos\theta_{t\gamma}
\end{equation}
in the center of
mass system and the prefactor $\kappa$ is given by 
\begin{equation}
  \kappa = 8\*\sqrt{2}\*\pi \aem\*\Gf\*\Vtb2\*\zw\*s.
\end{equation}
Here $\Gf$ denotes the Fermi constant
and $V_{tb}$ the Cabbibo-Kobayashi-Maskawa (CKM) matrix element.
The functions $l_1^{L},\, l_2^{L}$ are given by
\begin{eqnarray}
  l_1^{L}
  &=&
  2\*( -10\*\zb^3 -2\*\zb^2\*(1 - 11\*\zt) 
  + \zb\*(1 + 4\*\zt - 14\*\zt^2) 
  + \zt - 2\*\zt^2 + 2\*\zt^3 
  \nn\\&&
  -2\*\zw\*(19\*\zb^2 + \zt^2 +4\*\zb\*(1 - 3\*\zt))
  + \zw^2\*(\zt - 7\*\zb) ),\\
  &\nn\\
  l_2^{L}&=&
  8\*( \zb - 2\*\zb^2 + 2\*\zb^3 
  +  (1 + 4\*\zb - 8\*\zb^2)\*\zt 
  -2\*(1 - 5\*\zb)\*\zt^2 - 4\*\zt^3 
  \nn\\
  &&
  +  \zw\*(-2\*\zb^2 -2\*(4 - 9\*\zb)\*\zt - 20\*\zt^2) 
  +  \zw^2\*(\zb - 7\*\zt) ).
\end{eqnarray}
The above matrix element is given in 4 dimension. In the context
of the QCD corrections we will also need the squared matrix elements in
$d$ dimensions. The matrix element in $d=4-2\e$ dimensions 
for longitudinally polarized $W$-bosons 
is given by
\begin{equation}
  \sum_{\PolSum}|\cM_{0,d}^L|^2 = 
  \sum_{\PolSum}|\cM_0^L|^2
  +\e\*\kappa\*{2\*(\zb+\zt)\over\zw}\*\left[1-{1\over9}\*
  \left({2\over\xb-\ltb\*\cz}+{8\over\xt+\ltb\*\cz}\right)\right].
\end{equation}
In the derivation of the results above  we used
\begin{equation}
  \e_L^\mu(\kw) \e_L^\nu(\kw)^\ast =
  \L
  {\kw^\mu \kw^\nu \over \mw^2}
  -{\kw^\mu \kg^\nu + \kw^\nu \kg^\mu\over (\kw\cdot\kg)}
  + {\mw^2 \kg^\mu\kg^\nu\over  (\kw\cdot\kg)^2}
  \R
\end{equation}
with $\e_{L}(\kw)$ being the polarization vector of the incoming 
longitudinally polarized $W$ boson.
The squared matrix element for transversely polarized $W$-bosons reads:
\begin{eqnarray}
  \sum_{\PolSum}|\cM_0^T|^2
  &=&
  {2\*\kappa\over9\*\zw\*(1-\zw)^2}
  \*\Bigg(
  {64\*\zt\*\zw\*(\zb-\zt-\zw)\over\left(\xt+\ltb\*\cz\right)^2}
  -{16\*\zb\*\zw\*(\zb-\zt+\zw)\over\left(\xb-\ltb\*\cz\right)^2}
  \nn\\
  &+&
  {l_1^T\over\xb-\ltb\*\cz}
  +{l_2^T\over\xt+\ltb\*\cz}
  +6\*\cz\*\ltb\*\bigg(\zw\*(1-3\*\zb+3\*\zt)
  +6\*\zw^2\*(\zt-\zb)
  \nn\\
  &+&
  \zw^3\*(1+3\*\zb-3\*\zt)\bigg)
  -9\*\cz^2\*\ltb^2\*\zw\*(1+\zw^2)
  +\zw\*(-19-45\*\zb^2-18\*\zt
  \nn\\
  &-&
  45\*\zt^2+\zb\*(18+90\*\zt))
  +\zw^2\*(36\*\zb^2-\zb\*(12+72\*\zt)-60\*\zt+36\*\zt^2)
  \nn\\
  &+&
  \zw^3\*(-19-9\*\zb^2+6\*\zt-9\*\zt^2-\zb\*(6-18\*\zt))
  \Bigg)
\end{eqnarray}
with
\begin{eqnarray}
  l_1^{T}
  &=&
  4\*\zw\*\left(
    1 + 14\*\zb^2 + 2\*\zb\*(1 - 8\*\zt) - 2\*\zt + 2\*\zt^2
    + 2\*\zw\*(7\*\zb - \zt) + \zw^2\right), \\
  \nn\\
  l_2^{T}
  &=&
  16\*\zw\*\left(
    1 + 2\*\zb^2 - \zb\*(2+10\*\zt) + 2\*\zt + 8\*\zt^2
    - 2\*\zw\*(\zb - 4\*\zt) + \zw^2\right),
\end{eqnarray}
where we have used 
\begin{equation}
  \sum \e_T^\mu(\kw) \e_T^\nu(\kw)^\ast =
  \L -g^{\mu\nu} 
  + {\kw^\mu \kg^\nu + \kw^\nu \kg^\mu\over (\kw\cdot\kg)}   
  - {\mw^2 \kg^\mu\kg^\nu\over  (\kw\cdot\kg)^2}\R
\end{equation}
for the sum over the two different transverse polarizations
of the $W$-boson.

For later use we also give the squared matrix element for 
transversely polarized $W$-bosons in $d$ dimensions:
\begin{eqnarray} 
  &&\sum_{\PolSum}|\cM^T_{\rm d}|^2 = 
  \sum_{\PolSum}|\cM^T|^2
  +\e\*\kappa\*\Bigg[
  2\*\cz^2\*\ltb^2-{4\over3}\*\cz\*\ltb\*(1+3\*(\zb-zt))\nn\\
  &-&{1\over9\*(1-\zw)^2}\*\Bigg(
  {32\*\zb\*(\zb+\zt-\zw)\over(\xb-\ltb\*\cz)^2} 
  +{128\*\zt\*(\zb+\zt-\zw)\over(\xt+\ltb\*\cz)^2}\nn\\
  &+&{16\over\xb-\ltb\*\cz}
  \*\left(1-\zb-5\*\zb^2-(1+4\*\zb)\*\zt+\zt^2
    -(1-5\*\zb+\zt)\*\zw+\zw^2\right)\nn\\
  &+&{64\over\xt+\ltb\*\cz}\*\left(1-\zb+\zb^2
    -(1+\zb)\*\zt-2\*\zt^2-(1+\zb-2\*\zt)\*\zw+\zw^2\right)\nn\\
  &-&2\*\left(-36\*(\zb+\zt)+37\*(1+\zw^2)-38\*\zw
    +3\*(1-\zw)^2\*(2+3\*(\zb-\zt))\*(\zb-\zt)\right)
  \Bigg)\Bigg]\nn\\
  &-&4\*\kappa\*\e^2\*\left[1
    -{1\over9}\*\left({2\over\xb-\ltb\*\cz}+{8\over\xt+\ltb\*\cz}\right)\right]
  \label{eq:Mqt-d-dim}
\end{eqnarray}

Performing the remaining integration over the scattering angle
we obtain the leading-order total cross section $\sigtot^{T,L}$ for 
transversely/longitudinally polarized $W$-bosons:
\begin{equation}
  \sigtot^{T,L}=
  {1\over 9\* \sqrt{2}}\*
  {\aem\*\Gf\*\Vtb2\*N\over (1-\zw)^3} 
  \*{1\over \cN_{\,T,L}}\*
  \left( 
    l_1^{T,L}\* \ln\left({ 1+\beta_b\over 1-\beta_b}\right)
    + l_2^{T,L}\* \ln\left({ 1+\beta_t\over 1-\beta_t}\right)
    + K^{T,L} 
  \right),
  \label{eq:total-x-section}
\end{equation}
with
\begin{eqnarray}
  K^{L}&=&
  2\*\ltb\*\Big(
  28\*\zb^2 - 9\*\zt + 28\*\zt^2 - \zb\*(9 + 56\*\zt) 
  +2\*\zw\*(18\*\zb^2 + 41\*\zt + 18\*\zt^2 \nn\\
  &&+ \zb\*(5 - 36\*\zt))
  -\zw^2\*(-8 + 12\*\zb^2 -3\*\zb\*(1 + 8\*\zt) - 27\*\zt + 12\*\zt^2)
  \Big)\\
  K^{T}&=&
  4\*\zw\*\ltb\*\Big(
  -24\*\zb^2+6\*\zb\*(3 + 8\*\zt)
  -11-12\*\zt-24\*\zt^2 
  - 2\*\zw\*(-9\*\zb^2 + 3\*\zb\*(1+6\*\zt) \nn\\
  &&
  + 5 + 15\*\zt - 9\*\zt^2)
  -\zw^2\*(6\*\zb^2 - 12\*\zb\*\zt + 11 - 6\*\zt + 6\*\zt^2)
  \Big).
\end{eqnarray}
The velocities $\beta_t, \beta_b$ of the outgoing quarks in the 
center-of-mass system are given by
\begin{equation}
  \beta_t = {\ltb\over
    1+\zt-\zb},\quad\mbox{and}\quad
  \beta_b = {\ltb\over
    1+\zb-\zt}.
\end{equation}
Note that the relation between the cross sections for 
unpolarized, transversely
and  longitudinally polarized  $W$-bosons is given by
\begin{equation}
  \sigma = {1\over 3}(2\sigma^T+\sigma^L).
\end{equation}

Furthermore we note that the structure of the 
logarithmic terms   in \eq{eq:total-x-section} is universal and
can be obtained without an explicit calculation. In particular 
the singular contribution in the limit $\mb\to 0$ can be written as
\begin{equation}
  \sigma(W^+(\kw)\gamma(\kg)\to t(\kt) \bar b(\kb))
  \stackrel{m_b \to 0}{\longrightarrow} 
  \int_0^1 dx\; f_{b/\gamma}(x,\mb^2,s) \times
  \sigma(W^+(\kw)b(x\kg)\to t(\kt)),
\end{equation}
where $f_{b/\gamma}(x,\mb^2,s)$ can be interpreted as the bottom distribution
in the photon (at scale $s$):
\begin{equation}
  f_{b/\gamma}(x,\mb^2,s) = 
  {\alpha \over 2\pi} Q_b^2 P_{\gamma\to q\bar q}(x)
  \ln\L{s\over \mb^2}\R
\end{equation}
with the Altarelli-Parisi $P_{\gamma\to q\bar q}(x)$ kernel
\cite{AlPa77} given 
by 
\begin{equation}
P_{\gamma\to q\bar q}(x) = x^2+(1-x)^2.
\end{equation}
(We denote with $Q_b$ the electric charge of the $b$-quark in units
of the elementary charge $e$.)
A more detailed discussion will be given in section \ref{sec:results} where
the so-called structure function approach is investigated. 

\section{Virtual corrections}
\label{sec:virtual}
In this section we discuss the calculation of the virtual corrections.
In particular we sketch briefly a few technicalities of the calculation,
discuss  the ultraviolet (UV) and infrared singularities (IR), 
and carry out the renormalization.

We work in renormalized perturbation theory, which  means that the
bare quantities (fields and couplings) are expressed
in terms of renormalized quantities. By this procedure one obtains
two contributions: one is the original Lagrangian but now in terms of
the renormalized quantities, the second contribution are the so-called 
counter terms:
\begin{equation}
  \label{eq:ren-pert}
  {\cal L}(\Psi_0,A_0, m_0,g_0) = 
  {\cal L}(\Psi_R,A_R, m_R,g_R) + {\cal L}_{ct}(\Psi_R,A_R, m_R,g_R).
\end{equation}
The first contribution yields the same Feynman rules as the bare
Lagrangian but with the bare quantities replaced by renormalized ones.
In the following we renormalize the quark field and the quark mass 
in the on-shell scheme.
The conversion of the on-shell mass to the frequently used
\msbar\ mass or to any other renormalization scheme can be performed at the
end of the calculation. In spite of the fact that the calculation
presented here is a one-loop calculation, it is still leading-order in
the strong coupling constant $\as$. As a consequence the renormalization
of the coupling constant does not appear.

Whereas in the \msbar\ scheme the renormalization constants contain
only UV singularities, in the on-shell scheme they contain also
infrared  divergences.
We use dimensional regularization to treat both types of divergencies.
Although at the very end all the divergences must cancel
it is worthwhile to distinguish between UV and IR singularities so that
one can check the UV finiteness after renormalization and the
cancellation of the IR singularities independently.
\begin{figure}[htbp]
  \begin{center}
    \includegraphics[width=13cm,bb=29 340 559 642]{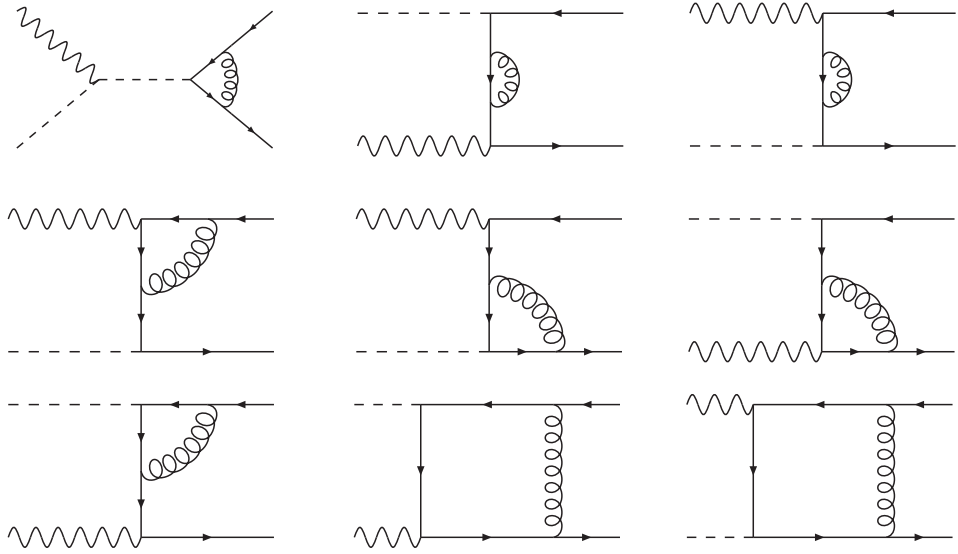}
    \caption{ Virtual corrections to $W^+ \gamma \to t \bar b$
      \label{fig:nlo}}
  \end{center}
\end{figure}
To start with let us discuss the contribution of the one-loop diagrams
before renormalization, that is the contribution from  
${\cal L}(\Psi_R,\ldots, m_R,g_R)$. The corresponding Feynman diagrams
are shown in fig.~\ref{fig:nlo}.
The calculation of the one-loop amplitude
is tedious but straightforward. To reduce the one-loop tensor 
integrals to scalar one-loop integrals we used the Passarino-Veltman 
reduction procedure \cite{PaVe79}.
The two one-loop box integrals  are
given by
\begin{eqnarray}
  D_{0,1}^{d} &=& {1\over i \pi^{2}} \int d^d\ell 
  {1\over (\ell^2-\mt^2)(\ell-\kt)^2 ((\ell-\kt-\kb)^2-\mb^2)
    ((\ell-\kt-\kb+\kg)^2-\mb^2)},\nn\\
  D_{0,2}^{d} &=& 
  \left.D_{0,1}^{d}\right|_{(\mt,\kt)\leftrightarrow (\mb,\kb)}.
\end{eqnarray}
All the simpler topologies follow from these integrals by dropping
one, two or three propagators. For example, using the standard notation
\cite{PaVe79} the infrared divergent triangle integral is given by
\begin{equation}
  C^d_0(1,2,3) = {1\over i \pi^{2}} \int d^d\ell 
  {1\over (\ell^2-\mt^2)(\ell-\kt)^2 ((\ell-\kt-\kb)^2-\mb^2)}.
\end{equation}
The one-point, two-point and the finite three-point integrals have been 
calculated in the standard way using Feynman parameterization. 
We have checked that our
results for these integrals agree with the numerical evaluation given 
by the FF package of G.~J.~van~Oldenborgh 
\cite{vanOldenborgh:1990wn,vanOldenborgh:1991yc}.
Here we give only explicit results for the two IR divergent integrals.
The triangle integral is given by
\begin{eqnarray}
  (2\pi\mu)^{2\e} \,\, \Re  C^d_0(1,2,3) &=&
  {1\over s\*\ltb}\*
  \Gamma(1+\e) \L {4\pi\mu^2\over s} \R^\e
  \*\bigg[{1\over\e}\*\ln(\rho)
  -{2\over3}\*\pi^2-2\*\ln(\ltb)\*\ln(\rho)
  \nn\\
  &-&{1\over4}\*(\ln(\wt)^2+\ln(\wb)^2)-\Li2(\wb)-\Li2(\wt)
  \bigg]+ O(\e)\nn\\
  &=&  {1\over s\*\ltb}\*
  \Gamma(1+\e) \L {4\pi\mu^2\over s} \R^\e
  {1\over\e}\*\ln(\rho)+\overline{C}_0(1,2,3),
\end{eqnarray}
with
\begin{equation}
  \omega_{i} = {1 - \beta_{i}\over 1 + \beta_{i} } \quad(i=t,b),
\end{equation}
and
\begin{equation}
  \label{eq:defRho}
  \rho 
  = \sqrt{{1-\zt-\zb - \ltb \over 1-\zt-\zb+ \ltb} }.
\end{equation}
To calculate the infrared divergent box integral we have used two
different methods. First we have considered a subtracted version of
the integral which can be calculated in 4 dimensions using Feynman
parameterization. From this result the desired result for the box
integral can be easily obtained.
The second method is based on the result given in ref.~\cite{BeDe90}. There 
the infrared singularity is regulated by a small photon mass
$\lambda$. This result can be converted to dimensional regularization
using the substitution \cite{MaSi75}
\begin{equation}
  \ln(\lambda^2) \to {1\over \e} - \gamma + \ln(4\pi\mu^2).
\end{equation}
We found
agreement of the results obtained by the two methods. The final expression
reads
\begin{eqnarray}
  &&(2\pi\mu)^{2\e} \Re D^d_{0,1} = 
  \*{1\over s^2\lambda_{tb}}
  {2\over (1-\zw)\* (\xb-\cz \ltb)}\*\Gamma(1+\e) \L {4\pi\mu^2\over s} \R^\e
  \nn\\
  && \times\bigg\{-{1\over \e}\ln(\rho)
  +
 \ln(\rho)\*\ln\L{{(\xb-\cz \*\ltb)^2\*\ltb^2\over 4\*\zb} }\R -\ln(\x3)^2
  \nn\\
  &&+2\*\ln\L{\x3+\rho\over 1+\rho\*\x3}\R \*\ln(\x3)
  -2\*\Li2(-\rho\*\x3)+\Li2(\rho^2)-2\*\Li2(-\rho/\x3)+{1 \over 2} \*\pi^2
  \bigg\},
\end{eqnarray}
with
\begin{equation}
  \x3= - {s \over 
     2\*\mb\*\mt}  \* (\zw-\zt-\zb+\lambda(\zw,\zb,\zt))
     =\sqrt{{\zw -\zt -\zb + \lambda(\zw,\zt,\zb) 
         \over \zw-\zt-\zb- \lambda(\zw,\zt,\zb)} }.
\end{equation}
In the actual calculation we have replaced the box integral in
$d=4-2\e$ dimensions by the box integral in $(d+2)$ dimensions. This
can be done by the use of the relation
\begin{equation}
  D_{27}^d = - {1\over 2\pi} D_0^{d+2}
\end{equation}
where $D_{27}^d$ is the coefficient of the metric tensor 
$g_{\mu\nu}$ in the decomposition
of the four-point tensor integral $D^d_{\mu\nu}$ 
(cf. eq. (F.3) in ref. \cite{PaVe79}),
which in turn can be expressed as a linear combination of the scalar box
integral $D_0^d$ and scalar triangle integrals $C_0^d(i,j,k)$ 
in $d$ dimensions.
This procedure has the advantage that only one infrared 
divergent integral appears. 
As a consequence the extraction of the IR divergent 
contribution is simplified. Very often this procedure yields also
a reduction of the algebraic complexity of the coefficients of the
scalar integrals. 
Using $D_0^{d+2}$ instead of $D_0^{d}$ we obtain the following result
for the contribution involving $C_0(1,2,3)$:  
\begin{eqnarray}
  \label{eq:IRdiv}
  \left.\delta|\cM^{T,L}_{1,d}|^2\right|_{\rm IR-div.} 
  = {\as\over \pi} \Cf  \*\left(\zb+\zt-1\right)\, 
  (2\pi\mu)^{2\e}\Re C_{0}(1,2,3)\,\,\*s\*
  \sum_{\PolSum}|\cM^{T,L}_{0,\rm d}|^2
\end{eqnarray}
(In the following discussion of the structure of the singularities we restrict 
ourselves  to the total cross section for unpolarized $W$-bosons.
The singular structure of the cross section for 
polarized $W$-bosons is analogous.)
As mentioned above the contribution involving $C_{0}^d(1,2,3)$
 is the only infrared divergent
contribution as far as the generic loop diagrams are concerned. 
Furthermore we note that no expansion in the dimensional
regulator $\e$ has been performed so
far. As a consequence we observe that the rational function multiplying
the $C^d_0(1,2,3) $  integral is up to an additional factor 
the squared born
matrix element in $d$ dimensions as it must be. This is an important
cross check of the calculation. The cancellation of the divergencies
by the real corrections is discussed in detail in section \ref{sec:real}.

Let us now switch to the UV divergent contribution which is generated
by the scalar one- and two-point integrals.
Defining the finite parts $\overline{A},\, \overline{B}$ of these integrals by
\begin{eqnarray}
  (2\pi\mu)^{2\e} A(\mt) &=& 
 {1\over \e} \Gamma(1+\e) \*
  \L{ 4\pi\mu^2\over \mt \mb }\R ^{\e}\* \mt^2 + \overline{A}(\mt),\nn\\ 
  (2\pi\mu)^{2\e} A(\mb) &=& 
 {1\over \e} \Gamma(1+\e) \*
  \L{ 4\pi\mu^2\over \mt \mb }\R ^{\e}\* \mb^2 + \overline{A}(\mb),\nn\\
 (2\pi\mu)^{2\e}B_0 &=& {1\over \e} \Gamma(1+\e)
  \L{ 4\pi\mu^2\over \mt \mb }\R^{\e} + \overline{B}_0,
\end{eqnarray}
the UV divergent contribution before renormalization reads:
\begin{eqnarray}
  \left.\delta|\cM_{1,d}|^2\right|_{\rm UV-div.} 
  &=&
  {\as\over 2\pi} \*\Gamma(1+\e)\Cf 
  \L{ 4\pi\mu^2\over \mt \mb }\R^{\e} {1\over \e} |\cM_0|^2 
  \nn\\
  &-&
  3\*{\as\over \pi } \*\Gamma(1+\e) \Cf 
  \L{ 4\pi\mu^2\over \mt \mb }\R^{\e}\*{1\over \e}\*F(\zw,\zb,\zt),
  \label{eq:UV-div-loop}
\end{eqnarray}
with
\begin{eqnarray*}
  &&F(\zw,\zb,\zt)={\kappa\over3\*\zw}\*\Bigg(
  z\*{\ltb\*(\zb-2\*\zt)}
  +{1\over2\*(1-\zw)\*\zw}\*(2\*\zb^2\*(1+\zw^2) 
  -\zb\*\big(1-5\*\zw-6\*\zw^2\nn\\
  &+&
  6\*\zt\*(1+\zw^2)\big) 
  -2\*\zt\*\big(1-5\*\zw-6\*\zw^2-2\*\zt\*(1+\zw^2)\big))\nn\\
  &+&
  {1\over3\*(1-\zw)^3\*\zw}\*\Bigg[
  \left\{{\zb^2\over(\xb-\ltb\*\cz)^3}
    +{4\*\zt^2\over(\xt+\ltb\*\cz)^3}\right\}\*
  64\*\zw\*(\zb^2-\zb\*(2\*\zt-\zw)+\zt^2\nn\\
  &+&
  \zt\*\zw-2\*\zw^2)
  -{8\*\zb\*\zw\over(\xb-\ltb\*\cz)^2}\*
  \Big\{10\*\zb^3+\zb^2\*(7-22\*\zt+5\*\zw) 
  -\zb\*\big(1-14\*\zt^2\nn\\
  &-&
  7\*\zw+24\*\zw^2+2\*\zt\*(5-7\*\zw)\big)
  +(1-\zt)\*\big(2\*\zt^2-2\*\zw\*(1+\zw)-\zt\*(1-3\*\zw)\big) \Big\}
  \nn\\
  &+& {32\*\zt\*\zw\over(\xt+\ltb\*\cz)^2}\*\Big\{
  2\*\zb^3 
  - \zb^2\*(3+8\*\zt-3\*\zw)
  + \zb\*\big(1+10\*\zt^2-5\*\zw-2\*\zw^2+2\*\zt\*(5-4\*\zw)\big)\nn\\
  &-&
  4\*\zt^3 
  - \zt^2\*(7-\zw)
  + \zt\*(1-7\*\zw+12\*\zw^2) 
  + 2\*\zw\*(1+\zw) \Big\}\\
  &-&{2\over(\xt+\ltb\*\cz)}\*\Big\{8\*\zb^4\*\zw 
  - 12\*\zb^3\*(1-\zw)\*\zw 
  - 2\*\zb^2\*\big(8\*\zt^2\*\zw-\zt\*(3-18\*\zw+31\*\zw^2) \nn\\
  &-& 2\*\zw\*(1-2\*\zw-5\*\zw^2) \big) 
  - \zb\*\big(4\*(1-5\*\zw)\*\zw^2 
  +4\*\zt^2\*(3-27\*\zw+16\*\zw^2) \nn\\
  &+&
  \zt\*(3-21\*\zw+41\*\zw^2+25\*\zw^3) \big)
  + \zt\*\big(8\*\zt^3\*\zw
  +6\*\zt^2\*(1-10\*\zw+9\*\zw^2)\nn\\ 
  &+& 4\*\zw\*(2+3\*\zw-6\*\zw^2+5\*\zw^3) 
  -
  3\*\zt\*(1-3\*\zw+19\*\zw^2-9\*\zw^3) \big) 
  \Big\}\\
  &-&{1\over(\xb-\ltb\*\cz)}\*\Big\{
  16\*\zb^4\*\zw 
  + \zb^3\*(6-72\*\zw+66\*\zw^2) 
  - \zb^2\*\big(3-21\*\zw+32\*\zt^2\*\zw+81\*\zw^2\nn\\
  &-&15\*\zw^3
  +4\*\zt\*(3-30\*\zw+11\*\zw^2)\big) 
+ \zb\*\big(\zt^2\*(6-24\*\zw+82\*\zw^2)+4\*\zw\*(1+3\*\zw+4\*\zw^3)\\
  &-&\zt\*(3-21\*\zw+73\*\zw^2+41\*\zw^3)\big)
-  8\*(1-\zt)\*\zt\*\zw\*\big(2\*\zt^2+\zw-5\*\zw^2-\zt\*(1-3\*\zw)\big) 
  \Big\}
    \Bigg]
       \Bigg).
\end{eqnarray*}
The UV singularities must be canceled by the renormalization procedure.
For the renormalization we need only to
consider the wave function renormalization $Z_\Psi$  of the quark
fields and the renormalization
of the mass parameters. As mentioned earlier the top quark mass and the bottom
mass are renormalized in the on-shell scheme. 
The generic counter term for a quark flavour  $f$ is given by
\begin{eqnarray}
  \label{eq:counter}
  \parbox[c]{3cm}{\includegraphics[bb=136 714 209 726]{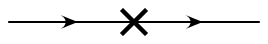}} &=& 
  i((Z_\Psi^\on(f)-1)\ksl -(Z_0^\on(f)-1)m_\on)\nn\\
  &=&  i \delta Z_\Psi^\on(f) (\ksl-m_\on) +i (\delta Z_\Psi^\on(f)- \delta
  Z_0^\on(f) ) m_\on,
\end{eqnarray}
with 
\begin{equation}
  Z_{\Psi,0}^\on(f) = 1 + \delta Z_{\Psi,0}^\on(f).
\end{equation}
The first term in \eq{eq:counter} gives simply the
corresponding born diagram multiplied by $- \delta Z_\Psi^\on$. 
This contribution itself is not gauge independent, it is canceled by
a similar contribution
from the $\gamma t\bar t$ and $\gamma b\bar b$ counter terms.
In addition to the mass counter term we have to consider the 
counter terms which corresponds 
to the vertex corrections. These counter terms amount to an additional
factor multiplying the born amplitude:
\begin{eqnarray}
  \parbox[c]{3.5cm}{\includegraphics[bb=111 681 208 762]{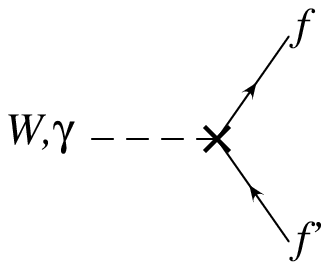}} &=&
  \parbox[c]{3.5cm}{\includegraphics[bb=111 681 208 762]{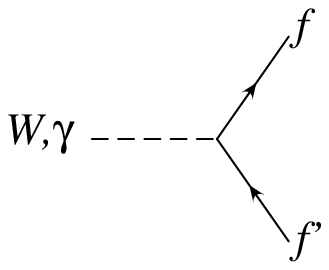}}
  \times\; (\sqrt{Z_\Psi^\on(f)}\sqrt{Z_\Psi^\on(f')} - 1),
\end{eqnarray}
where $f (f')$ denote the flavours of the outgoing quark 
(anti-quark).
 Expanding the factor in the coupling we obtain 
\begin{equation}
  \sqrt{Z_\Psi^\on(f)}\sqrt{Z_\Psi^\on(f')} - 1
  = {1\over 2} ( \delta Z_\Psi^\on(f)
  + \delta Z_\Psi^\on(f') ).
\end{equation}
The contribution from the renormalization is thus given by
\begin{equation}
  \cM_1^{\st{ren}} =  {1\over 2} ( \delta Z_\Psi^\on(t)
  +\delta Z^\on_\Psi(b)) 
  \cM_{0,\rm d} + \cM_1^{\st{ct}}
\end{equation}
where $\cM_1^{\st{ct}}$ denotes the contribution of the term
$i (\delta Z_\Psi- \delta Z_0 )m_\on$ in \eq{eq:counter}.
The contribution to the squared matrix element finally reads
\begin{equation}
  \cM_1^{\st{ren}}{\cM_{0,\rm d}}^\ast
  + {\cM_1^{\st{ren}}}^\ast\cM_{0,\rm d}
  = ( \delta Z_\Psi(t)+\delta Z_\Psi(b))|\cM_{0,\rm d}|^2
  + 2 \Re (\cM_1^{\st{ct}}{\cM_{0,\rm d}}^\ast).
\end{equation}
Using 
\begin{eqnarray}
  \delta Z_\Psi^{\st{on}}(f) &=& 
  {\as\over 4\*\pi}\*\Gamma(1+\e)\*\Cf  
  \L{4\*\pi\*\mu^2\over m_f^2}\R ^{\e} \* \left\{
    -{1\over \e}
    +{2\over \epsIR} 
    - 4 \right\}+ O(\e) \nn\\
  \delta Z_\Psi^{\st{on}}(f)- \delta Z_0^{\st{on}}(f) 
  &=&
  {\as\over 4\*\pi}\Gamma(1+\e)\*\Cf  \L {4\pi\mu^2\over m_f^2}\R ^{\e}
  \left\{  {3\over \e} + 4 \right\} + O(\e).
  \label{eq:ren-const}
\end{eqnarray}
we obtain 
\begin{eqnarray}
  &&  \cM_1^{\st{ren}}{\cM_{0,\rm d}}^\ast
  + {\cM_1^{\st{ren}}}^\ast\cM_{0,\rm d} =\nn\\
  &&\hspace{-1cm} {\as\over 4\*\pi}\*\Gamma(1+\e)\*\Cf  
  \* \bigg\{
  -{2\over \e}\L{4\*\pi\*\mu^2\over \mt\mb}\R ^{\e}
  +
  {2\over \epsIR} \L{4\*\pi\*\mu^2\over \mt^2}\R ^{\e}
  +{2\over \epsIR} \L{4\*\pi\*\mu^2\over \mb^2}\R ^{\e}
  - 8 \bigg\} |\cM_{0,\rm d}|^2\nn\\
  &&\hspace{-1cm}
  +  {\as\over \pi}\Gamma(1+\e)\*\Cf  
  \left\{  {3\over \e} + 4 \right\} 
  \bigg(\L {4\pi\mu^2\over \mt^2}\R ^{\e} f_t(\zw,\zb,\zt)
  + \L {4\pi\mu^2\over \mb^2}\R ^{\e}f_b(\zw,\zb,\zt)
  \bigg)
  \label{eq:UV-div-ren}
\end{eqnarray}
In \eq{eq:ren-const} we have introduced $\epsIR$ (with $d=4+2\epsIR$)
 to distinguish
between IR and UV singularities. 
The functions $f_t, f_b$ are given in the appendix.
Inspecting \eq{eq:UV-div-ren}
one observes that the UV singularities match exactly those appearing
in \eq{eq:UV-div-loop}. The UV singularities are thus canceled by 
the renormalization procedure as it should be. Note that the
renormalization procedure introduces an additional IR divergent
contribution
\begin{eqnarray}
  {\as\over 2\*\pi}\*\Gamma(1+\e)\*\Cf   {1\over \epsIR}
  \L
  \L{4\*\pi\*\mu^2\over \mt^2}\R ^{\e} 
  +
  \L{4\*\pi\*\mu^2\over \mb^2}\R ^{\e} 
  \R \* 
   |\cM_{0,\rm d}|^2.
\end{eqnarray}
The complete IR divergent contribution of the virtual corrections is
thus given by: 
\begin{eqnarray}
  &&
  {\as\over \pi} 
  \Gamma(1+\e) \L {4\pi\mu^2\over s} \R^\e
  \Cf  
  {\left(\zb+\zt-1\right)\over \ltb}\*
  \*\ln(\rho)
  {1\over\e}
  \sum_{\PolSum}|\cM_{0,\rm d}|^2
  \nn\\&&
  - {\as\over 2\*\pi}\*\Gamma(1+\e)\*\Cf   
  \L
  \L{4\*\pi\*\mu^2\over \mt^2}\R ^{\e} 
  +
  \L{4\*\pi\*\mu^2\over \mb^2}\R ^{\e} 
  \R \* {1\over \e}
   \sum_{\PolSum} |\cM_{0,\rm d}|^2.
  \label{eq:IRdivVirtual}
\end{eqnarray}
The cancellation of the IR singularities is discussed in the next section.
\section{Real corrections}
\label{sec:real}
In this section we consider the calculation of the real corrections
\begin{equation}
  W^+(\kw) + \gamma(\kg)  \to t(\kt) + \bar b(\kb) + g(\k1).
\end{equation}
The calculation of the matrix elements is straightforward and does
not impose any problems. In principle it can also be done
automatically with packages like for example CompHEP \cite{Pukhov:1999gg} or
MadGraph \cite{Stelzer:1994ta}.
We have checked that we reproduce in the soft limit the factorization
formulae
\begin{equation}
  \cM_0(\kg,\kw,\kt,\kb,\k1) 
  \stackrel{\k1 {\mbox{\scriptsize soft}}}{\longrightarrow}
  \cS(\kt,\k1,\kb) \times \cM_0(\kg,\kw,\kt,\kb)
\end{equation}
with the well known eikonal factor
\begin{eqnarray}
  \cS(k_i,k_s,k_j) =
  {2\* (k_i\cdot k_j)\over (k_i\cdot k_s)\*(k_j\cdot k_s) }
  -{(k_i\cdot k_i)\over (k_i\cdot k_s)^2}
  -{(k_j\cdot k_j)\over (k_j\cdot k_s)^2}.
\end{eqnarray}
We also compared the results numerically with MadGraph and found agreement.
The IR divergencies arise from the phase space integration over
regions where the gluon is soft. To extract these singularities we
used the subtraction method for massive quarks 
\cite{Phaf:2001gc,Catani:2002hc}. The basic
idea of the subtraction method is to add and subtract a so-called dipole term
which on the one hand matches point-wise the singularities in the 
real corrections and on the other hand is simple enough to allow an 
analytic integration over the unresolved phase space in $d$
dimensions \cite{Catani:2002hc}:
\begin{eqnarray}
  \sigma^{\st{NLO}} &=&
  \int dR(\kt,\kb) \bigg[d\sigma^{\st{virt.}}(\kg,\kw,\kt,\kb) 
  + d\sigma^{\st{born}}(\kg,\kw,\kt,\kb) 
  \otimes {\bf I}\bigg]_{\e=0}\nn\\
  &+& 
  \int dR(\kt,\kb,\k1) \bigg[ 
  \left.d\sigma^{\st{real}}(\kg,\kw,\kt,\kb,\k1)\right|_{\e=0}\nn\\
  && - \sum_{\st{dipoles}} d\sigma^{\st{born}}(\kg,\kw,\kt,\kb) 
  \otimes dV_{\st{dipoles}}\bigg|_{\e=0}
  \bigg],
  \label{eq:master}
\end{eqnarray}
In \eq{eq:master} the symbol $\otimes$ involves in addition 
to the identification of the kinematics also spin correlations. In
general also colour correlations appear. This is not the case here
because the leading-order matrix element is proportional to the unit
matrix in colour space.
We note that the dipoles
$dV_{\st{dipoles}}$ are universal and can be obtained from the study
of soft and collinear limits \cite{Catani:2002hc}. 
The dependence on the specific process
is encoded in $d\sigma^{\st{born}}(\kg,\kw,\kt,\kb)$.
The integral of the dipoles over the `dipole phase space' which
appears in the factorization of the phase space is denoted by
\begin{equation}
   {\bf I} =  \int_1 \sum_{\st{dipoles}}dV_{\st{dipoles}}.
\end{equation}
A more detailed description of the subtraction method is given in
ref.~\cite{Catani:2002hc}, 
here we just reproduce  the relevant formulae for the
specific case studied in this paper. 
In the notation of ref. \cite{Catani:2002hc} the dipoles
$dV_{\st{dipoles}}$  are given by  
\begin{eqnarray}
  \sum_{\st{dipoles}} d\sigma^{\st{born}}(\kg,\kw,\kt,\kb) 
  \otimes dV_{\st{dipoles}}&=&{1\over 2\*(\k1\cdot\kt)}
  \* \bra V_{1t,b}\ket |\cM_0(\tilde k_{1t},\tilde\kb)|^2\nn\\
  &+& {1\over 2\*(\k1\cdot\kb)}
  \* \bra V_{1b,t}\ket |\cM_0(\tilde\kt,\tilde k_{1b})|^2
\end{eqnarray}
with
\begin{equation}
  \bra V_{1t,b}\ket = 8\*\pi\*\as\*\Cf\*
  \L {2\over 1-\tilde z_t \* (1-y_{1t, b})} 
  - {\tilde v_{1t, b} \over v_{1t, b}}\* \bigg[1+\tilde z_t +
  {\mt^2\over (\kt\cdot{\k1}) }\bigg] \R.
\end{equation}
The momenta $\tilde k_{ij}, \tilde k_k$ play the role of the emitter
and the spectator. For the detailed definition we refer to ref.
\cite{Catani:2002hc}.
The general expressions for $\tilde z_i,y_{ij, k}, v_{ij, k},\tilde v_{ij, k}$ 
are also given in ref. \cite{Catani:2002hc}.
For the specific reaction considered here we obtain
\begin{equation}
  \tilde z_t = {(\kt\cdot\kb)\over (\kt\cdot\kb) + ({\k1}\cdot\kb) },
\end{equation}
\begin{equation}
  y_{1t, b} = { (\kt\cdot{\k1} )\over
    (\kt\cdot{\k1}) + ({\k1}\cdot\kb ) + (\kt\cdot\kb ) },
\end{equation}
\begin{equation}
  \tilde v_{1t, b} = {\ltb
    \over 1-\zt-\zb},
\end{equation}
and
\begin{equation}
  v_{1t, b} = {\sqrt{[2\*\zb+(1-\zt-\zb)\*(1-y_{1t,b})]^2
      -4\*\zb}\over (1-\zt-\zb)\*(1-y_{1t,b})}.
\end{equation}
Combining the dipole terms as given in \eq{eq:master}\  together with 
the real corrections given by $d\sigma^{\st{real}}(\kg,\kw,\kt,\kb,\k1)$ the
integration can be done numerically in 4 dimensions over the whole phase space.
The integrals of the dipoles over the dipole phase space (which we
have to add to account for the additional term we have subtracted from
the real corrections) can be obtained from ref. \cite{Catani:2002hc}: 
\begin{equation}
  {\bf I} = 
  {\as\over 2\pi}{1\over \Gamma(1-\e)} 
  \L{4\pi\mu^2\over s}\R^\e
  \Cf [ 4\,\* I^{\eik}_+(\sqrt{\zt},\sqrt{\zb};\e)
  +I^{\coll}_{1t,b}(\sqrt{\zt},\sqrt{\zb};\e)
  +I^{\coll}_{1b,t}(\sqrt{\zb},\sqrt{\zt};\e)
  ]
\end{equation}
with
\begin{eqnarray}
  I^{\eik}_+(\mu_j, \mu_k;\e) &=&
  {1\over \tilde{v}_{ij,k}}\*
  \Bigg[ {1\over 2\*\e} \* \ln(\rho)
  - \ln(\rho)\,\*\ln\Big(1 - (\mu_j + \mu_k)^2\Big)
  - {1\over 2}\* \ln(\rho_j(\mu_j, \mu_k))^2 
  \nn \\ && 
  - {1\over 2}\* \ln(\rho_k(\mu_j, \mu_k))^2
  +{\pi^2\over 6}
  + 2\* \Li2(-\rho) - 2\* \Li2(1 - \rho)
  \nn \\ && 
  - {1\over 2}\* \Li2(1 - \rho_j^2(\mu_j, \mu_k)) 
  - {1\over 2}\* \Li2(1 - \rho_k^2(\mu_j, \mu_k))\Bigg]
  + O(\e){} \nn\\
  &\equiv&
  \frac{1}{\tilde{v}_{ij,k}}
  \frac{1}{2\e} \ln \rho + \hat I^{\eik}_+(\mu_j, \mu_k),
\end{eqnarray}
\begin{eqnarray}
  I^{\coll}_{gQ,k}(\mu_Q,\mu_k;\e) & =&
  \frac{1}{\e}+\ln(\mu_Q)
  -2\*\ln\left[(1-\mu_k)^2-\mu_Q^2\right]+\ln(1-\mu_k)
  \nn\\
  && 
  -{2\*\mu_Q^2 \over 1-\mu_Q^2-\mu_k^2}\*\ln\left({\mu_Q\over 1-\mu_k}\right)
  +5-{\mu_k\over 1-\mu_k}-{2\*\mu_k\*(1-2\*\mu_k)\over 1-\mu_Q^2-\mu_k^2}
  +O(\e){}\:,\nn\\
  &\equiv&
  \frac{1}{\e}+ \hat I^{\coll}_{gQ,k}(\mu_Q,\mu_k),
\end{eqnarray}
and
\begin{equation}
  \rho_n(\mu_j, \mu_k) =
  \sqrt{{1-\tilde{v}_{ij,k}+2\mu_n^2/(1-\mu_j^2-\mu_k^2)\over
      1+\tilde{v}_{ij,k}+2\mu_n^2/(1-\mu_j^2-\mu_k^2)}}, \quad(n=j,k)
  \mbox{ with }   \mu_i = \sqrt{z_i}
\end{equation}
and $\rho$ as defined in \eq{eq:defRho}.
From the formulae above we can read off the singular contribution
\begin{equation}
  {\as\over 2\pi}{1\over \Gamma(1-\e)} 
  \L{4\pi\mu^2\over s}\R^\e
  \Cf \frac{2}{\e} \left( \frac{1-\zt-\zb}{\ltb}
    \ln(\rho) +  1\right) |\cM_{0,\rm d}|^2
\end{equation}
Comparing the above result with \eq{eq:IRdivVirtual} we observe that 
the real corrections indeed cancel the IR divergent contribution from
the virtual corrections. Having canceled the IR divergencies all the
remaining phase space integrals can now be done numerically in 4
dimensions.

\section{Results}
\label{sec:results}
Before presenting  the results we first discuss a few consistency
checks. As mentioned earlier we have checked the loop integrals
appearing in the virtual corrections with the 
FF package of G.~J.~van~Oldenborgh 
\cite{vanOldenborgh:1990wn,vanOldenborgh:1991yc} or in the case of
the box integral by comparison with results
available in the literature. Using the box integrals in $d+2$
dimensions we have verified that only the triangle integral
$C_0^d(1,2,3)$ produces an IR singularity and that the form of this
singularity agrees with the structure predicted by QCD. By this procedure
we test essentially the coefficients of the two box integrals and the
IR divergent triangle integral. Note that this check is valid in $d$
dimensions. That means that the $\gamma$-algebra is also tested in 
$d$ dimensions. (The treatment of $\gamma_5$ is not an issue here because we
have only four external momenta.)
Furthermore we have checked that the UV singularities have
exactly the form as predicted by the renormalization procedure. The 
structure of the UV singularities is determined by the one- and
two-point integrals. The coefficients of these
integrals (more precisely a linear combination of them) are thus
checked 
 by the
fact that we reproduce the predicted structure of the UV singularities.
We have checked the real corrections by the comparison with Madgraph. A
further important check is also the finiteness of the real corrections
in combination with the subtraction terms discussed in the previous
section. This is a non-trivial 
check because the matrix elements are tested point wise in the
singular regions. 

Let us now come to the numerical results. 
For the numerical evaluation we have chosen the following parameters: 
\begin{equation}
  \Gf =1.16639\times 10^{-5}\mbox{GeV}^{-2}, \quad
  \alpha={1\over 137.036},\quad 
  \mw=80.42~\mbox{GeV}, \quad
  \mt =175~\mbox{GeV}.
\end{equation}
For the strong coupling we have used a running $\alpha_s$ with the 
renormalization scale set to  the center of mass energy. 
As input value we used $\alpha_s(\mu=200\mbox{ GeV}) = 0.105$. 
Note that the Fermi constant $\Gf$ and the electric coupling $\alpha$ 
enter only through a prefactor and
can thus be changed without redoing the numerical integration.
For the $b$-quark pole mass we consider the range between $4.6$ and
$5.1$ GeV as given by the particle  data group \cite{pdg2002}.   
\begin{figure}
  \includegraphics[width=0.49\textwidth,bb=22 35 545 525]{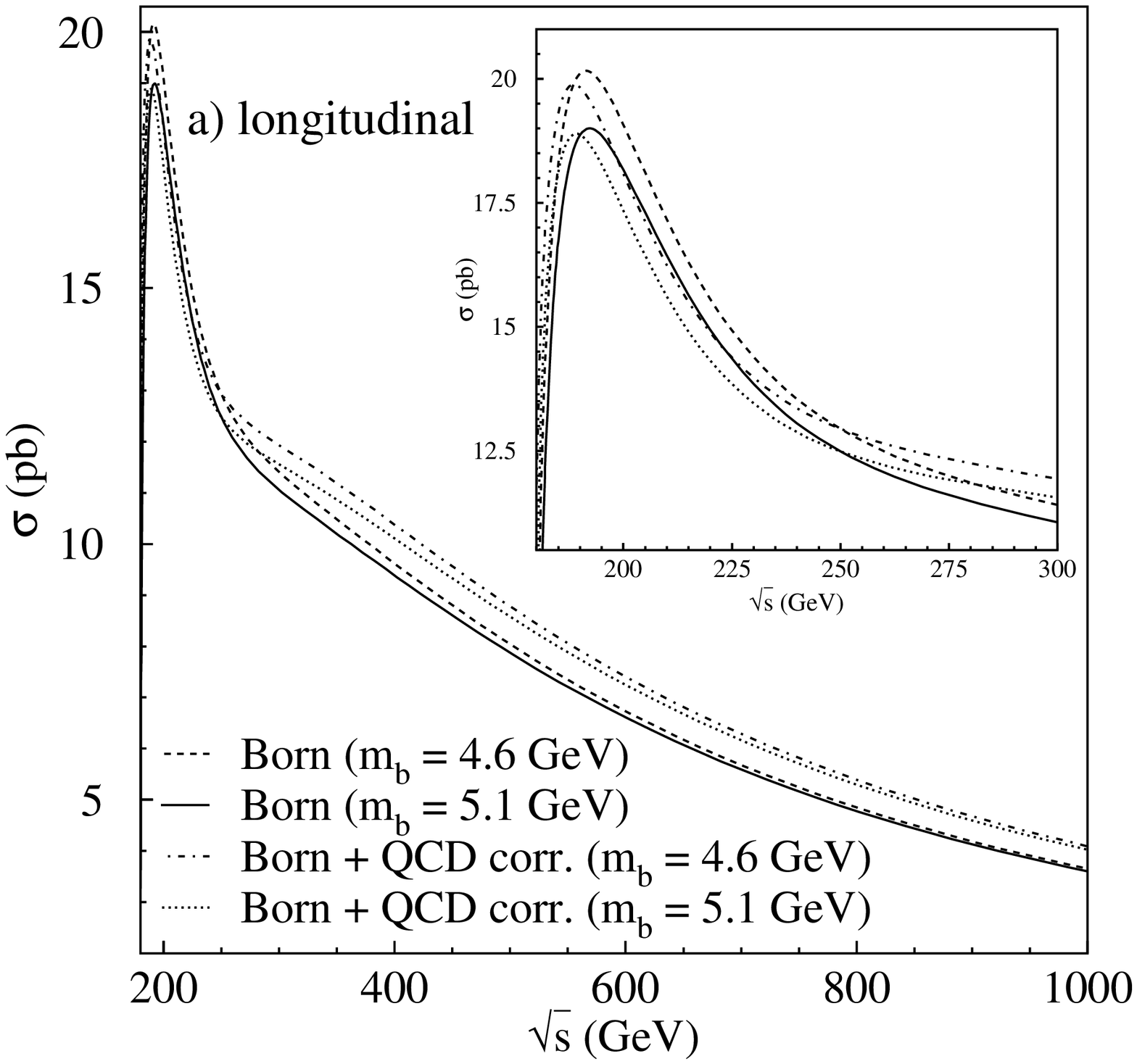}    
  \includegraphics[width=0.49\textwidth,bb=22 35 545 525]{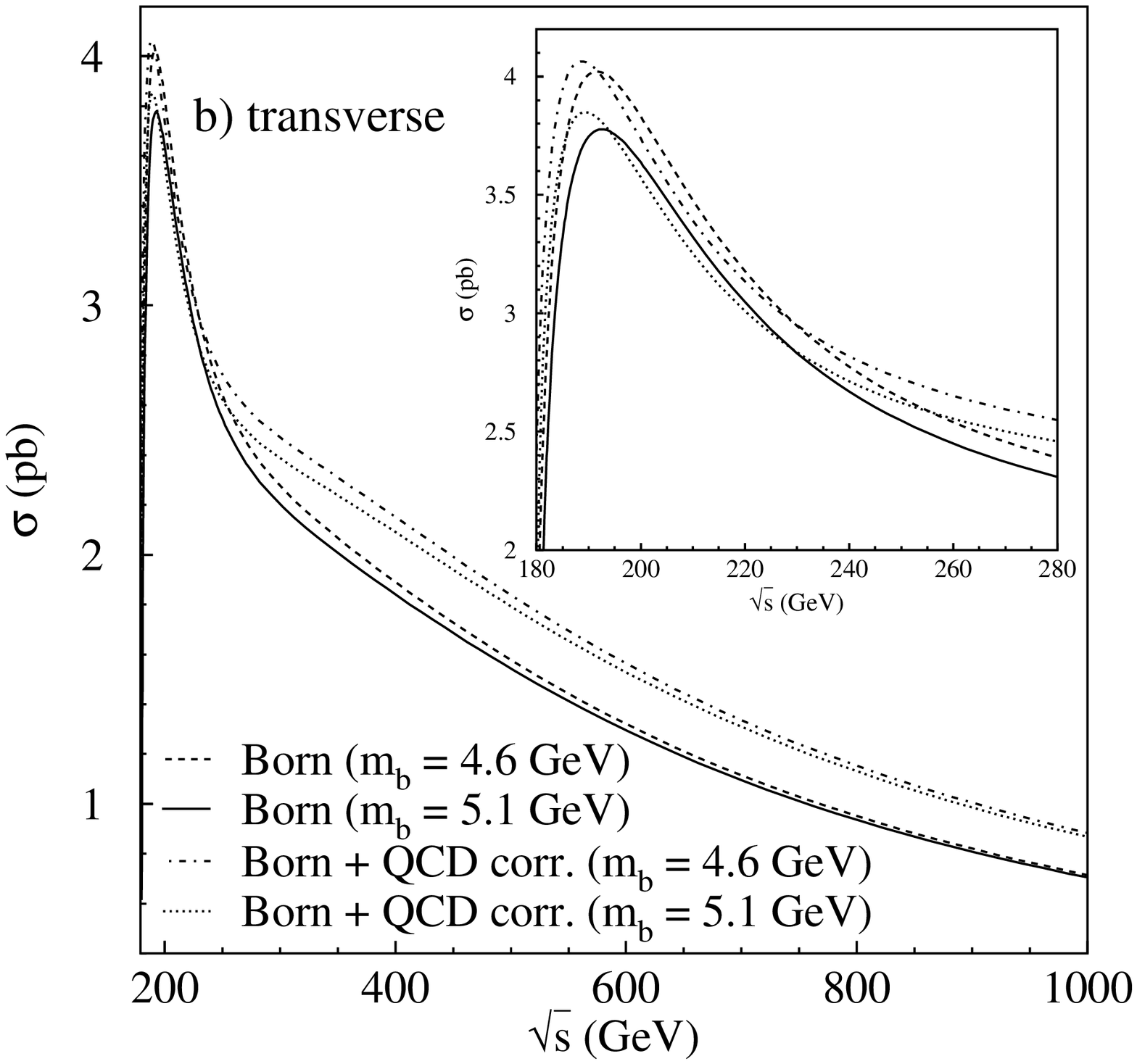}
  \caption{\label{crosstot}
    Total cross section for the process 
    $W^+\gamma\rightarrow t\bar{b}$ 
    for polarised W-Bosons. 
    The smaller figures inside the plots show the threshold region.
    }
\end{figure}
\begin{figure}
  \begin{center}
    \parbox{\textwidth}{
    \includegraphics[width=0.49\textwidth, bb=27 29 414 488]{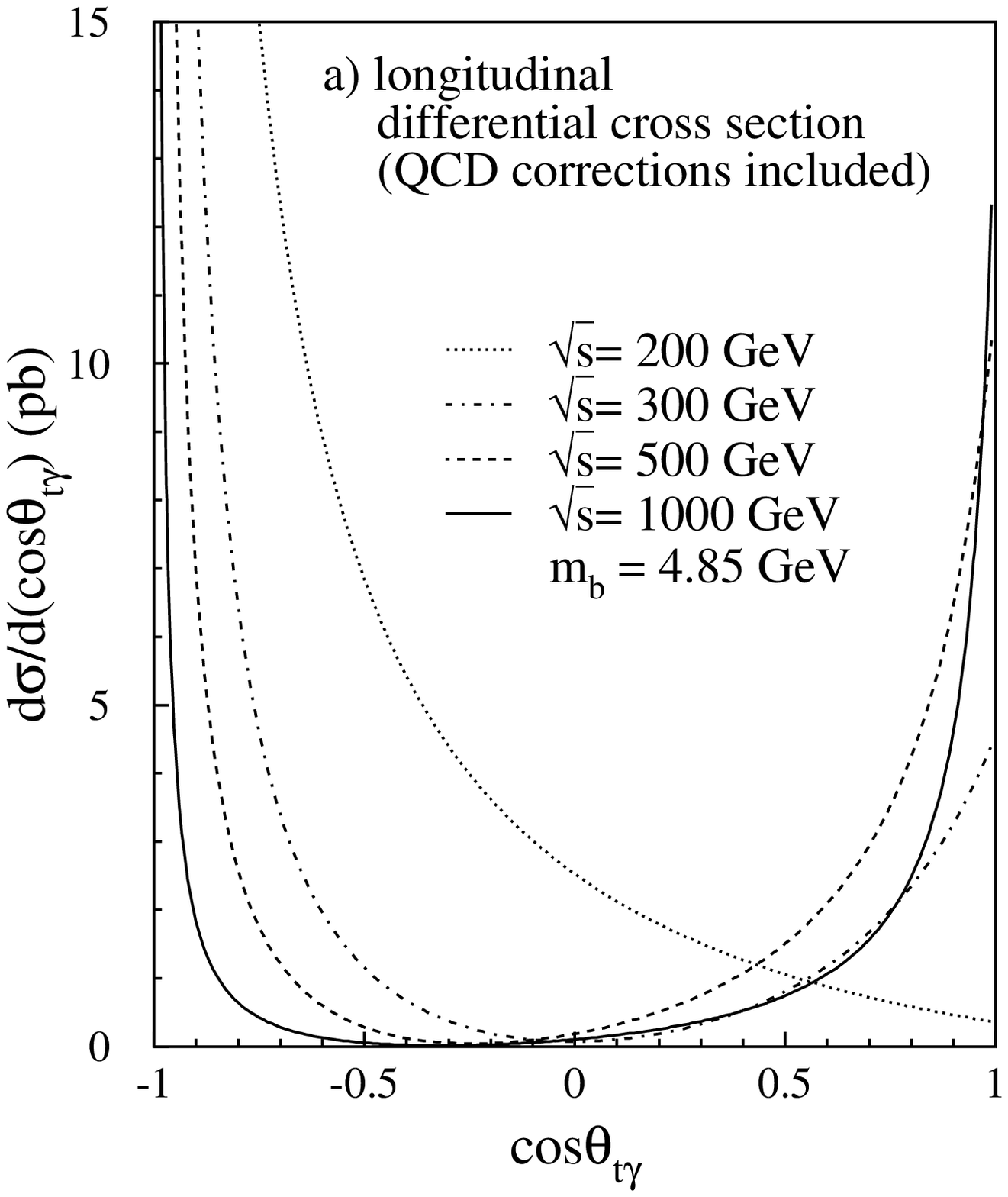}    
    \includegraphics[width=0.49\textwidth, bb=27 29 414 488]{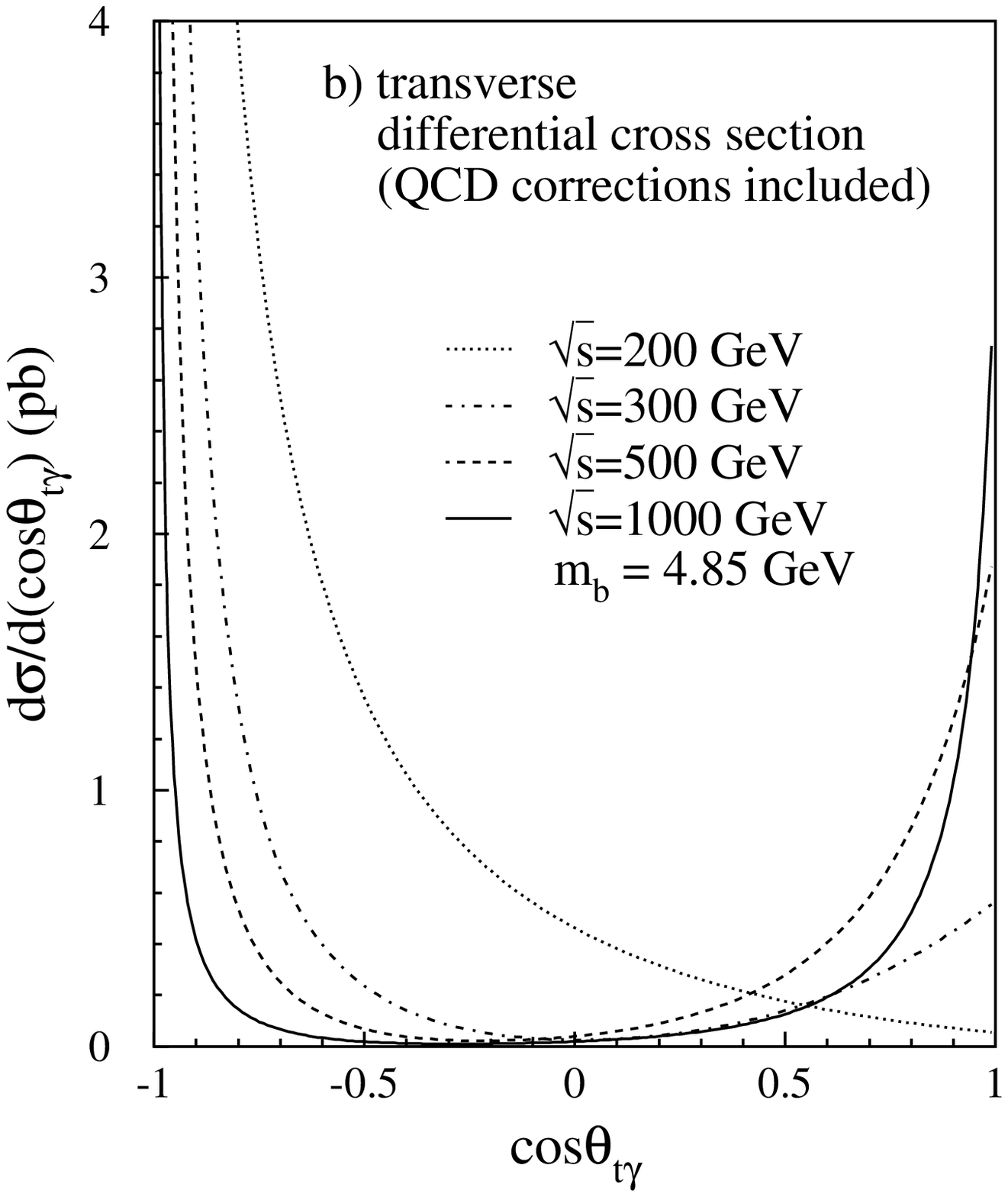}}
  \end{center}
  \caption{\label{diffcross}
    Differential cross section for the process 
    $W^+\gamma\rightarrow t\bar{b}$ 
    with polarised W-Bosons 
    as a function of $\cos(\theta_{t\gamma})$}
\end{figure}
In \fig{crosstot} the total cross section for the process 
$W^+\gamma\rightarrow t\bar{b}$ is shown for polarised $W$-bosons.
Fig. \ref{crosstot}a shows the total cross section for longitudinally polarised
$W$-bosons, whereas fig. \ref{crosstot}b is the corresponding 
plot for the transversely
polarised case. Both plots show the Born cross section as well as 
the QCD corrected cross sections for two different values of 
the $b$-quark mass. The QCD corrections are of 
the order of 12\% in the longitudinally polarised case 
and of the order of 24\% in the transversely
polarised case for a center of mass energy of 1000 GeV.
In the smaller figures inside the two plots the threshold region is
shown. Here one observes that the QCD corrections become negative for a
center of mass energy between 195 and 225 GeV. Close to the threshold
the corrections are again positive. As one might expect from
phase space arguments the cross section for a
$b$-quark mass of 4.6 GeV is larger than the cross section for a $b$-quark
mass of 5.1 GeV. We note that the difference between the two different
mass values is quite sizeable in the energy range 300--600 GeV
given the smallness of $b$-quark mass compared to the center of
mass energy. The relative size in this region is roughly given by
$\ln(m_{b1}^2/m_{b2}^2)/\ln(m_{b1}^2/s)$. For a center of mass energy
arround 500 GeV this corresponds to an effect of around 2.2\%.
For larger center of mass energies the curves approach rapidly. 
Furthermore we note that the cross section for transversely
polarised $W$-bosons is suppressed in comparison with the one for
longitudinal $W$-bosons. This is a well known feature of the $W$-boson
couplings to very massive quarks. In 
ref. \cite{Kunszt:1988tk}
it has been demonstrated that in the space-like axial gauge with 
a specific parmetrization of the Higgs-doublet the contribution of
longitudinally polarized gauge bosons comes primarily from the
`scalar gauge fields'. In particular in this specific gauge the 
equivalence theorem
\cite{Chanowitz:1985hj,Lee:1977eg,Cornwall:1974km,Vayonakis:1976vz} 
known from the  $R_\xi$-gauge 
becomes an identity in the sense of an expansion in $\mw^2\over s$. 

Let us add at this point also a remark about the renormalization scale
uncertainty. As told already the QCD corrections are of leading-order
in the strong coupling constant. That means no compensation of the 
residual scale dependence takes place. If we consider for example 
for a $\mu$ value of 500 GeV the range between 250 and 1000 GeV 
we obtain a variation of the QCD corrections by about 7-8 \%. Keeping in
mind that the QCD corrections are only of the order of 10-20 \% this
yield an uncertainty of the cross section of 1.5 \%. We can thus
conclude that the scale dependence is small as far as the total cross
section is concerned.

In \fig{diffcross} the differential cross section is shown as a
function of $\cos\theta_{t\gamma}$, with  $\theta_{t\gamma}$ defining
the angle between the initial state photon and the final state
top quark. Figure a) shows again the longitudinally polarised case 
and figure b) the transversely polarised one. In the curves shown
the QCD corrections are included. Both cross sections increase strongly for
the case that the angle $\theta_{t\gamma}$ becomes close to $180$ degrees.
The origin of this behaviour is an initial state collinear singularity
which appears for massless $b$-quarks, when the initial state photon and
the final state $b$-quark become collinear. In the case of a 
none-vanishing  $b$-quark mass this singularity is regulated by the 
finite $b$-quarks mass and becomes manifest as a $\ln(\mb^2/s)$. 
When the photon and the top quark become collinear the increase of the
cross section is smaller since the top quark mass is not so small
compared to the center of mass energies considered in fig.~\ref{diffcross}.\\
In principle these logarithmic terms
could be large and one might worry that the convergence of
the perturbative expansion is spoiled. As far as QED is concerned 
this is here not the case as long as one considers only moderate values
for the center of mass energy. Then $\alpha \ln({\mb^2\over
  s})$ is still a small quantity 
(i.e. $~{1\over 137}\* \ln({5^2\over 1000^2})=-0.077$) and
perturbation theory remains applicable. 
Although these logarithms are not
a problem at leading-order it is worthwhile to study their resummation. 
This is interesting in itself for two reasons. On the one
hand the framework to do so is the so-called structure function
approach for massless quarks (only the $b$-quark is considered massless) --- 
which, in principle, one could have used from the begining. In the
structure function approach the logarithms are absorbed into the
structure functions and resummed via an Altarelli-Parisi like
evolution. As mentioned earlier the QED evolution itself is not important
for moderate values of the center of mass energy. 
Therefore one might argue that
the structure function approach for massless $b$-quarks should give a good
description because terms of order ${\mb^2\over s}$ 
--- which are dropped in this approach --- are small. In general this
is not true because close to the threshold region the $b$-quark mass 
effects can be important. 
It is therefore instructive to compare the fixed-order
calculation with the structure function approach. The second reason
why the structure function approach is of interest are the
logarithms appearing in the QCD corrections. In principle they could
be larger and the need for the resummation becomes more
important\footnote{%
From a practical oriented viewpoint those logarithms if present 
can not cause a serious problem because otherwise the QCD
correction  would be much
larger. Nevertheless one should address this issue in the future to get a 
more reliable prediction.}. The
theoretical framework would be once again the structure function
approach but now with a mixed evolution. 

In the structure function approach the total cross section for top quark
production reads:
\begin{eqnarray}
  \sigma &=& \int dx\, \Gamma_{\gamma/\gamma}(\mu_F,x) 
  \times \hat \sigma(W^+(\kw)\gamma(x\kg)\to t(\kt)\bar b(\kb))\nn\\
  &+&  \int dx\, \Gamma_{b/\gamma}(\mu_F,x) 
  \times \hat \sigma(W^+(\kw)b(x\kg)\to t(\kt))
\end{eqnarray}
The cross sections appearing in the above equation are the subtracted
cross sections for massless $b$-quarks. They depend on the
factorization scheme used to factorize the singular part.
We used the \msbar\  scheme. Note that although
not written down explicitly the cross sections depend in general also on the
factorization scale $\mu_F$ at which the subtraction is done. The
functions $\Gamma_{\gamma/\gamma}(\mu_F,x)$,
$\Gamma_{b/\gamma}(\mu_F,x)$ are the `parton distribution functions' 
describing the probability to find a photon or a $b$-quark inside a
photon. The above procedure is in fact almost the same as the corresponding
procedure in QCD describing hadron-hadron reactions. However there is
one important difference: in QCD the structure functions are
not calculable in perturbation theory, while  in QED it is possible to
calculate the structure functions perturbatively.
To the order needed here they are given by
\begin{eqnarray}
  \Gamma_{\gamma/\gamma}(\mu_F,x) & = & \delta(1-x)+O(\alpha),\nn\\
  \Gamma_{b/\gamma}(\mu_F,x) & = & 
  {\alpha\over 2\pi} Q_b^2 
  (x^2+(1-x)^2)
  \*\ln \left({\mu_F^2 \over \mb^2 }\right)+O(\alpha^2).
  \label{eq:SF-simple}
\end{eqnarray}
Note that the $\Gamma_{\gamma/\gamma}(\mu_F,x)$ structure function is
only needed to order $\alpha^0$ because the subtracted hard scattering
cross section starts already with $\alpha$. The above results for the
structure functions can be
easily obtained from a matching calculation. In fact one could also argue
that one starts with the initial conditions 
$ \Gamma_{\gamma/\gamma}(\mb,x)= \delta(1-x)$, 
$\Gamma_{b/\gamma}(\mb,x)  = 0$ and generates the above distributions
dynamically through evolution. This gives the same result.

In the following discussion we restrict ourselves to the unpolarized 
cross section, the polarized case can be discussed in the same way.
Using 
\begin{eqnarray}
  &&\hat \sigma(W^+(\kw)\gamma(\kg)\to t(\kt)\bar b(\kb)) = \nn\\
  &&
  {1 \over 27}\* {1\over \sqrt{2}}\*\alpha\* \Gf \*  |V_{tb}|^2\* N\*
  {1\over (1-\zw)^3}\*
  \bigg(
  (1-\zt)\*( -2\*\zw\*(11+6\*\zw+11\*\zw^2)
  \nn\\
  &&
  -(9-58\*\zw+33\*\zw^2-12\*\zw^3)\*\zt
  +4\*(7-3\*\zw^3+6\*\zw^2-3\*\zw)\*\zt^2
  ) 
  \nn\\&&
  -4 \*(\zt+2\*\zw)
  \*(1-2\*\zt+4\*\zw\*\zt+\zw^2-4\*\zt^2)
  \* \ln(\zt)\nn\\
  &&
  + (\zt+2\*\zw)
  \*(1-2\*\zt-2\*\zw\*\zt+\zw^2+2\*\zt^2)
  \*\ln\left({(1-\zt)^2\*s\over
    \mu_F^2 }\right)
  \bigg)
\end{eqnarray}
for the \msbar\  subtracted parton cross section
we obtain the following result for the leading-order cross section
in the structure function approach: 
\begin{eqnarray}
  \sigma 
  &=&
  {1 \over 27}\*{1\over \sqrt{2}}\*\alpha\* \Gf\*  |V_{tb}|^2\*N\* 
  {1\over (1-\zw)^3}\*
  \bigg((1-\zt)\*(
  -2\*\zw\*(11+6\*\zw+11\*\zw^2)\nn\\
  &&-(9-58\*\zw+33\*\zw^2-12\*\zw^3)\*\zt
  + 4\*(7-3\*\zw+6\*\zw^2-3\*\zw^3)\*\zt^2
  )
  \nn\\&&
  -4 \*(\zt+2\*\zw)
  \*(1-2\*\zt+4\*\zw\*\zt+\zw^2-4\*\zt^2)
  \* \ln(\zt)\nn\\&&
  +
  (\zt+2\*\zw)
  \*(1-2\*\zt-2\*\zw\*\zt+\zw^2+2\*\zt^2)
  \*\ln\left({(1-\zt)^2\over
    \zb }\right)
  \bigg)
\end{eqnarray}
Note that we have used the structure functions given in
\eq{eq:SF-simple} --- and not the evolved ones --- 
which is strictly speaking only valid for 
$\mu_F\approx \mb$. As mentioned earlier limiting ourselves to
center of mass energies up to 1 TeV the evolution does not change
the structure functions very much.
\begin{figure}
  \begin{center}
\includegraphics[width=0.65\textwidth,bb=40 224 469 625]{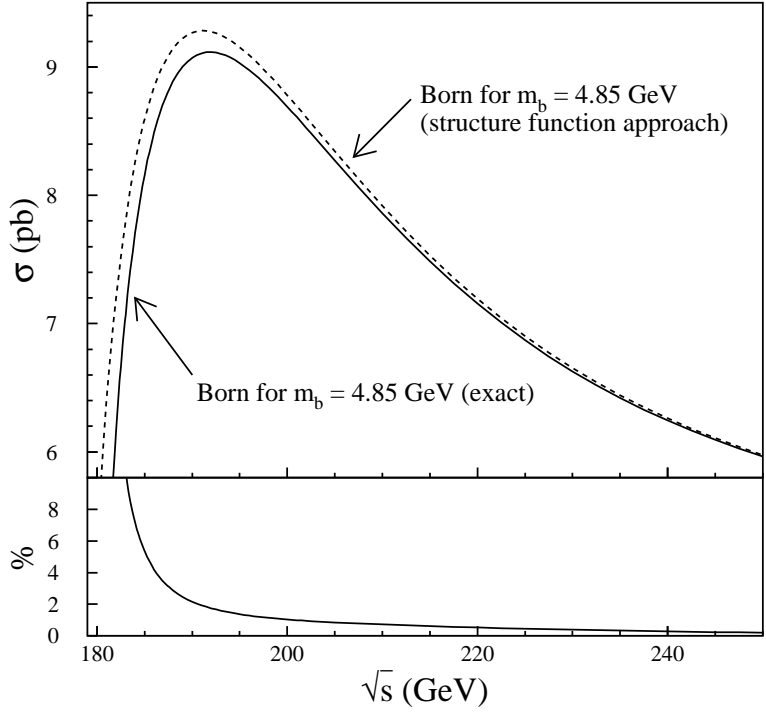}    
  \end{center}
  \caption{
    \label{fig:distr}
    Total born cross section for the process 
    $W^+\gamma\rightarrow t\bar{b}$
    for unpolarized W-Bosons.
    The upper figure compares the structure function approach with 
    the fixed order calculation for massive $b$-quarks. 
    The lower figure shows the deviation of the 
    structure function approach from the massive calculation
    in percent.}
\end{figure}   
We have checked that  the difference which one obtains using on the
one hand the evolved structure functions and on the other hand
the fixed order results (\eq{eq:SF-simple}) 
is indeed of the order of a few per mil and thus  negligible.
Keeping only terms involving $\ln(\mb)$ in the exact result 
\eq{eq:total-x-section}, and
dropping all terms which vanish in the limit $\mb\to 0$, we reproduce
the above result.
The comparison between the two approaches is shown in \fig{fig:distr}.
It is clearly visible that for large center of mass energies the
structure function approach agrees with the fixed order calculation. 
This is due to the fact that the corrections of the type $\mb^2/s$ can be
neglected at high energies and that the logarithms of the form
$\ln({\mb^2\over s})$ are still of moderate size so that the resummation of
these terms would not change the result. In the threshold region 
one observes a significant difference between the two methods. In this
region the finite $b$-quark mass is important because it affects the
location of the threshold. 

So far we have only studied the reaction $W^+\gamma \to t\bar b$ which 
is not directly observable. While a high energy photon `beam' can be realized
through the interaction of low energy photons of a laser with
high energy electrons/positrons (see for example
ref. \cite{Kuhn:1993fx}) a $W$-boson beam is
not  available.
On the other hand one can argue that the dominant contribution to
the reaction 
\begin{equation}
  \ell^+\gamma \to t\bar b \bar \nu_\ell \quad (\ell = e,\mu)
\end{equation}
proceeds via the production of an almost on-shell $W$-boson which 
then interacts with
the photon to produce the top quark and the $b$-quark. This is 
the essence of the so called {\it effective $W$-boson approximation} 
\cite{Kane:1984bb,Dawson:1985gx,Lindfors:1985yp,Kunszt:1988tk,Kauffman:1989ca}.
In this approach the intermediate $W$-bosons is considered on-shell
and described through structure functions similar to the afore
mentioned structure function approach for the $b$-quark.
The effective $W$-approximation is similar to the
Weizsäcker-Williams \cite{Weiszacker1934} approximation.
The total cross section for $\ell^+\gamma \to t\bar b \bar \nu_\ell$ 
in this approach is then given by 
\cite{Kane:1984bb,Dawson:1985gx,Lindfors:1985yp,Kunszt:1988tk,Kauffman:1989ca}
\begin{equation}
  \sigma(\ell^+\gamma\rightarrow t\bar{b}\bar{\nu}_e)
  = 
  \int\!dx\,\,f_{W_{L}^+/\ell^+}(x)\,\,
  \,\sigma(W_{L}^+\gamma\rightarrow t\bar{b})  
  +
  \int\!dx\,\,f_{W_{T}^+/\ell^+}(x)\,\,
  \,\sigma(W_{T}^+\gamma\rightarrow t\bar{b}),  
\end{equation}
with the structure functions $f_{W_{T,L}}$ given by
\begin{equation}
  f_{W_L/\ell}(x)=
  {\alpha\over4\,\pi\,\sin\theta_{w}^{2}}\,{1-x\over x}\,,
\end{equation}
\begin{equation}
  f_{W_T/\ell}(x)=
  {\alpha\over8\,\pi\,\sin\theta_{w}^{2}}\,{1+(1-x)^2
    \over x}\,\ln\left({s\over m_{w}^2}\right)\,.
\end{equation}
Note that we have written down only the leading terms for the structure
functions. The `sub-leading' terms which are suppressed by additional
powers of $\mw^2/s$ are not universal and depend on the exact
prescription how to define them. We note that the distribution function
of longitudinally polarized $W$-bosons is very well approximated by
the leading term. On the other hand using only the leading term for
the structure function of transversly polarized $W$-bosons results in 
an overestimate of the cross section for energies of the order of 1
TeV. For the structure function $ f_{W_T/\ell}(x)$ we have included
the sub-leading terms as given in ref. \cite{Dawson:1985gx}.
An additional remark on the use of those functions is in order: while 
in the original work \cite{Dawson:1985gx} a lower boundary on the
allowed $x$ values appears ($x>\mw/E$), apparently no such boundary appears in 
refs. \cite{Dawson:1987tc,Kunszt:1988tk}. Having studied the quality 
of the approximation for center of mass energies of about 40 TeV 
and heavy $t,b$-quarks we find that only without this additional
constraint we obtain good agreement. For the present case we have 
used the following approach: for the contribution of the longitudinally 
polarized $W$-bosons the constraint $x>\mw/E$ is not used. For the
transversly polarized $W$-bosons we must use the additional
constraint because otherwise the distribution function (including
${\mw^2\over s}$ corrections, see eq. (2.18) in
ref. \cite{Dawson:1985gx})  is not defined.
\begin{figure}
  \begin{center}
      \includegraphics[width=0.8\textwidth,bb=16 35 546 348]{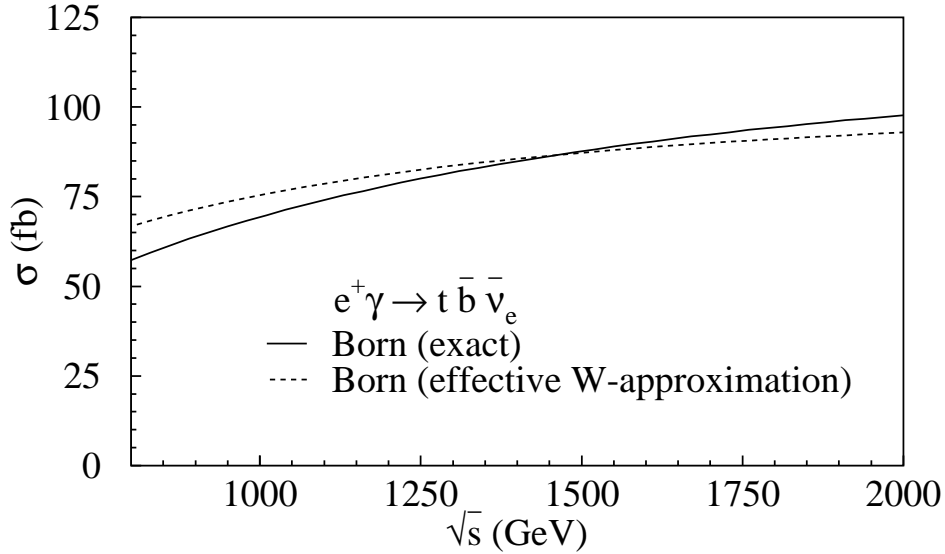} 
  \end{center}
  \caption{\label{fig:Wapprox-vs-exact}
    Total cross section for the process $e^+\gamma\rightarrow
    t\bar{b}\bar{\nu}_e$. The full line shows the exact 
    result while the dashed line shows the result using 
    the effective $W$-approximation. 
    }
\end{figure}
In \fig{fig:Wapprox-vs-exact} 
we show the leading order result for the reaction 
$\ell^+\gamma \to t\bar b \bar \nu_\ell$. The full line is the exact
result. The exact result  agrees with the result presented in 
ref. \cite{Boos:2001sj}. 
The dashed line shows the result in the effective
$W$-approximation using the above prescription.

\begin{figure}
  \begin{center}
    \includegraphics[width=0.8\textwidth,bb=16 35 546 348]{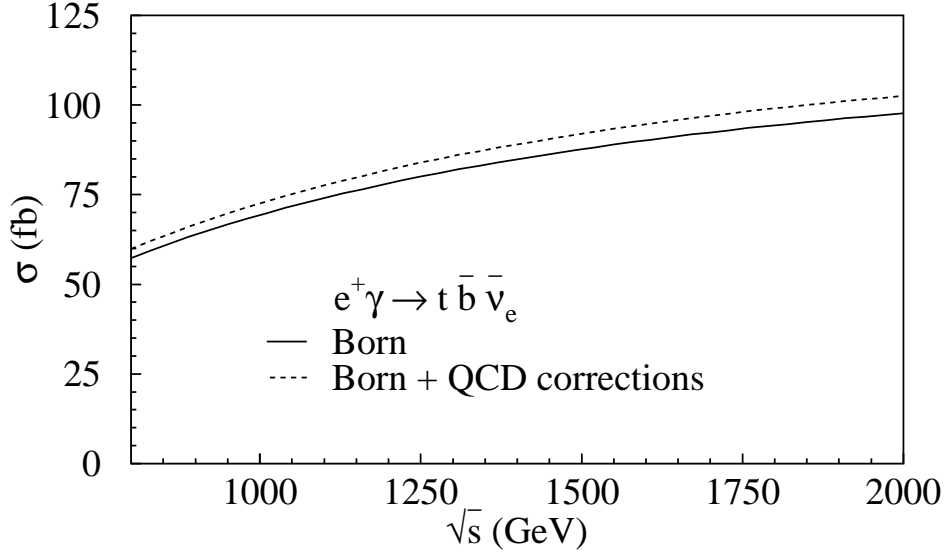} 
  \end{center}
  \caption{\label{fig:egtb}
    Total cross section for the process $e^+\gamma\rightarrow
    t\bar{b}\bar{\nu}_e$ in the effective $W$-approximation. 
    }
\end{figure}

It is clearly visible that the accuracy of the approximation is only at 
the 10\% level for small values of the center of mass energy. 
To obtain a reliable prediction at NLO we have
combined the exact leading-order result with the QCD corrections
obtained in the effective $W$-approximation. We expect that by this
procedure the uncertainty due to the  effective $W$-approximation is
smaller than a percent and thus of the same order as
the next-to-next-to-leading order QCD corrections. The NLO cross
section obtained by this procedure is shown in \fig{fig:egtb}.
We note that the QCD corrections which are quite sizeable at the level
of the partonic reaction $W^+\gamma  \to t\bar b$ for large values of
the center of mass energy are only of the order of 5\% for the reaction 
$e^+\gamma \to t\bar b \bar \nu_e$. This is due to the convolution
with the $W$-distribution functions which gives more weight to the
lower center of mass energy values.

\section{Conclusions}
\label{sec:conclusions}
In the present paper we have studied the QCD corrections for 
single top quark production in electron photon interactions. We have
first calculated the QCD corrections for $W^+\gamma\to t\bar b$. 
Applying the effective $W$-approximation these results can be
used to obtain the QCD corrections for $e^+\gamma\to t\bar b \bar
\nu_e$. While the corrections are sizeable for the reaction  
$W^+\gamma\to t\bar b$ they are only of the order of  5\% for the 
reaction $e^+\gamma\to t\bar b \bar\nu_e$. We can thus conclude that 
as far as the QCD corrections are concerned the reaction 
$e^+\gamma\to t\bar b \bar\nu_e$ is very well suited for precise
measurements of the CKM matrix element $V_{tb}$.

{\bf Acknowledgments:}\\
One of us (P.U.) would like to thank Arnd Brandenburg,
Sven Moch, Markus Roth and 
Stefan Weinzierl for useful discussions, and Arnd Brandenburg for a
careful reading of the manuscript.
C.S. would like to thank the Land Baden-Württemberg 
and the Graduiertenkolleg for High Energy Physics and Particle
Astrophysics for financial support. This work has been supported by
BMBF contract 05HT1VKA3.

\appendix
\section{Contributions from self-energy like counter terms}
\renewcommand\theequation{\ifnum \value{section}>0
 \Alph{section}.\arabic{equation}%
\else
\arabic{equation}%
\fi}

\begin{eqnarray}
f_t(\zw,\zb,\zt)&=&
{\kappa\over3\*\zw}\*\Bigg(
{\zt\over\zw\*(1-\zw)}\*
\left\{2\*\zt\*(1+\zw^2)-2\*\zb\*(1+\zw^2)+6\*\zw^2+5\*\zw-1\right\}\nn\\
&+&{256\*\zt^2\over3\*(1-\zw)^3\*(\xt+\ltb\*z)^3}\*
\left\{\zb^2-\zb\*(2\*\zt-\zw)+\zt^2+\zt\*\zw-2\*\zw^2\right\}\nn\\
&+&{32\*\zt\over3\*(1-\zw)^3\*(\xt+\ltb\*z)^2}\*
\Big\{2\*\zb^3-\zb^2\*(3+8\*\zt-3\*\zw) 
+\zb\*(1+10\*\zt^2\nn\\&+&\zt\*(10-8\*\zw)-5\*\zw-2\*\zw^2)
-4\*\zt^3-\zt^2\*(7-\zw)+\zt\*(1-7\*\zw\nn\\&+&12\*\zw^2)+2\*\zw+2\*\zw^2
\Big\}
-{2\*\zt\over3\*\zw\*(1-\zw)^3\*(\xt+\ltb\*z)}\*
\Big\{8\*\zb^3\*\zw+\zb^2\*(6\nn\\&-&44\*\zw-8\*\zt\*\zw+38\*\zw^2)
-\zb\*(3-17\*\zw+8\*\zt^2\*\zw+17\*\zw^2+21\*\zw^3\nn\\
&+&\zt\*(12-104\*\zw+60\*\zw^2))+8\*\zt^3\*\zw
+\zt^2\*(6-60\*\zw+54\*\zw^2)\nn\\
&-&\zt\*(3-9\*\zw+57\*\zw^2-27\*\zw^3)+20\*\zw^4-24\*\zw^3
+12\*\zw^2+8\*\zw \Big\}\nn\\
&-&{8\*\zt\over3\*(1-\zw)^3\*(\xb-\ltb\*z)}\*
\Big\{2\*\zb^3 + \zb^2\*(1-2\*\zt-\zw) 
+\zb\*(1-2\*\zt^2-6\*\zw\nn\\&-&\zw^2+\zt\*(2+6\*\zw))
+2\*\zt^3-3\*\zt^2\*(1-\zw)+\zt\*(1-2\*\zw-5\*\zw^2)\nn\\&-&\zw+5\*\zw^2\Big\}
-2\*z\*\zt\*\ltb
-\e\*\bigg[
{4\*\zt\*(1+2\*\zw)\over(1-\zw)}
+{512\*\zw\*\zt^2\*(\zt+\zb-\zw)\over3\*(1-\zw)^3\*(\xt+\ltb\*z)^3}\nn\\
&+&{32\*\zt\over3\*(1-\zw)^3\*(\xt+\ltb\*z)^2}\*
\Big\{4\*\zb^2\*\zw
-\zb\*(4\*\zt\*\zw+\zw^2+8\*\zw-1) 
-8\*\zt^2\*\zw\nn\\&+&\zt\*(15\*\zw^2-12\*\zw+1)
+2\*\zw^3-2\*\zw^2+4\*\zw\Big\}\nn\\
&-&{8\*\zt\over3\*(1-\zw)^3\*(\xt+\ltb\*z)}\*
\Big\{4\*\zb^2\*\zw+\zb\*(8\*\zt\*\zw-5\*\zw^2-2\*\zw-1) 
+4\*\zt^2\*\zw\nn\\
&+&\zt\*(11\*\zw^2-22\*\zw+3)
+11\*\zw^3-16\*\zw^2+7\*\zw+2\Big\}\nn\\
&-&{16\*\zt\over3\*(1-\zw)^3\*(\xb-\ltb\*z)}\*
\Big\{2\*\zb^2\*\zw
+\zb\*(4\*\zt\*\zw-\zw^2-4\*\zw+1) \nn\\
&+&2\*\zt^2\*\zw-2\*\zt\*\zw\*(1+\zw)+2\*\zw^2 \Big\}
\bigg]
\Bigg)
+O(\e^2)
\end{eqnarray}

\begin{eqnarray}
f_b(\zw,\zb,\zt)&=&
{\kappa\over3\*\zw}\*\Bigg(
{\zb\over2\*\zw\*(1-\zw)}\*
\left\{2\*\zb\*(1+\zw^2)-2\*\zt\*(1+\zw^2)+6\*\zw^2+5\*\zw-1\right\}\nn\\
&+&{64\*\zb^2\over3\*(1-\zw)^3\*(\xb-\ltb\*z)^3}\*
\left\{\zb^2-\zb\*(2\*\zt-\zw)+\zt^2+\zt\*\zw-2\*\zw^2\right\}\nn\\
&-&{8\*\zb\over3\*(1-\zw)^3\*(\xb-\ltb\*z)^2}\*
\Big\{10\*\zb^3+\zb^2\*(7-22\*\zt+5\*\zw)
+\zb\*(14\*\zt^2\nn\\&+&\zt\*(14\*\zw-10)-24\*\zw^2+7\*\zw-1)
-2\*\zt^3+3\*zt^2\*(1-\zw)+\zt\*(2\*\zw^2\nn\\&+&5\*\zw-1)-2\*\zw^2-2\*\zw
\Big\}
-{\zb\over3\*\zw\*(1-\zw)^3\*(\xb-\ltb\*z)}\*
\Big\{16\*\zb^3\*\zw-\zb^2\*(16\*\zt\*\zw\nn\\&-&66\*\zw^2+72\*\zw-6)
-\zb\*(16\*\zt^2\*\zw
     +\zt\*(36\*\zw^2-112\*\zw+12)-15\*\zw^3\nn\\&+&81\*\zw^2-21\*\zw+3)
+16\*\zt^3\*\zw+\zt^2\*(34\*\zw^2-40\*\zw+6)
-\zt\*(3-13\*\zw\nn\\&+&25\*\zw^2+33\*\zw^3)+16\*\zw^4+12\*\zw^2+4\*\zw\Big\}
-{8\*\zb\over3\*(1-\zw)^3\*(\xt+\ltb\*z)}\*
\Big\{2\*\zb^3\nn\\&-&\zb^2\*(2\*\zt-3\*\zw+3)
-\zb\*(2\*\zt^2-2\*\zt\*(1+3\*\zw)+5\*\zw^2+2\*\zw-1)
+2\*\zt^3\nn\\&-&\zt^2\*(\zw-1)-\zt\*(\zw^2+6\*\zw-1)+5\*\zw^2-\zw \Big\}
+z\*\zb\*\ltb\nn\\
&-&\e\*\bigg[
{2\*\zb\*(1+2\*\zw)\over(1-\zw)}
+{128\*\zw\*\zb^2\*(\zb+\zt-\zw)\over3\*(1-\zw)^3\*(\xb-\ltb\*z)^3}\nn\\
&-&{8\*\zb\over3\*(1-\zw)^3\*(\xb-\ltb\*z)^2}\*
\Big\{20\*\zb^2\*\zw+\zb\*(16\*\zt\*\zw-27\*\zw^2+12\*\zw-1 )\nn\\
&-&4\*\zt^2\*\zw+\zt\*(\zw^2+8\*\zw-1) 
-2\*\zw^3+2\*\zw^2-4\*\zw \Big\}\nn\\
&-&{4\*\zb\over3\*(1-\zw)^3\*(\xb-\ltb\*z)}\*
\Big\{ 8\*\zb^2\*\zw + \zb\*(16\*\zt\*\zw+7\*\zw^2-26\*\zw+3)
+ 8\*\zt^2\*\zw\nn\\&-& \zt\*(7\*\zw^2+10\*\zw-1)
+10\*\zw^3-11\*\zw^2+8\*\zw+1\Big\}\nn\\
&-&{16\*\zb\over3\*(1-\zw)^3\*(\xt+\ltb\*z)}\*
\Big\{2\*\zb^2\*\zw + \zb\*(4\*\zt\*\zw-2\*\zw^2-2\*\zw) 
+ 2\*\zt^2\*\zw \nn\\&-& \zt\*(\zw^2+4\*\zw-1)
+ 2\*\zw^2\Big\}
\bigg]
\Bigg)
+O(\e^2)
\end{eqnarray}


\newcommand{\zp}{Z. Phys. }\def\as{\alpha_s }\newcommand{\prd}{Phys. Rev.
  }\newcommand{\pr}{Phys. Rev. }\newcommand{\prl}{Phys. Rev. Lett.
  }\newcommand{\npb}{Nucl. Phys. }\newcommand{\psnp}{Nucl. Phys. B (Proc.
  Suppl.) }\newcommand{\pl}{Phys. Lett. }\newcommand{\ap}{Ann. Phys.
  }\newcommand{\cmp}{Commun. Math. Phys. }\newcommand{\prep}{Phys. Rep.
  }\newcommand{\jmp}{J. Math. Phys. }\newcommand{\rmp}{Rev. Mod. Phys. }

\end{document}